\documentclass[12pt]{article} 
\usepackage{graphicx}
\usepackage{amsmath}
\usepackage[varg]{txfonts}
\begin{document}
\title{Mirror mesons at the Large Hadron Collider (LHC)}
\author{George Triantaphyllou \\~\\
National Technical University of Athens\\
            58 SINA Str., GR 106 72 ATHENS, Greece, \\
   email: gtriantaphyllou@aya.yale.edu}
\maketitle
\noindent {\bf Abstract} \\
The existence of mirror partners (katoptrons) of Standard-Model (SM) fermions offers a viable alternative to
a fundamental BEH mechanism, with the coupling corresponding to
 the gauged mirror generation symmetry becoming naturally strong at energies around 1 TeV.
The resulting 
non-perturbative processes produce dynamical katoptron masses which might
range from 0.1 to 1.15 TeV in a way circumventing 
usual problems with the $S$ parameter. 
Moreover, they create
mirror mesons belonging in two main groups, with masses differing from each other approximately by a factor of six and which might range  approximately from 0.1 to 2.8 TeV. Since 
the corresponding phenomenology expected at hadron colliders is particularly rich,
 some interesting mirror-meson cross-sections are
  presented, something that might also lead to a deeper understanding of the underlying mirror fermion structure.
Among other findings, results in principle
 compatible with indications from LHC concerning decays of new particles to two photons are 
analyzed.

\section{Motivation}
Higher luminosities and collision energies of proton beams at CERN have recently raised hopes that a new structure 
behind the BEH mechanism will be revealed shortly. In the present work, 
efforts are focused on studying the 
phenomenological implications of the existence of strongly-interacting mirror fermions at energies accessible at
the LHC.  Several models describing a strongly-interacting electroweak sector have been developed over the 
last decades \cite{Hill}. However, speculation on the existence of mirror fermions first appeared in \cite{LeeYang} and having
them constitute an effective, dynamical electroweak BEH mechanism \cite{Weinb} appeared in \cite{WilZee}.
The gauged katoptron-generation symmetry is expected to confine these mesons
 so that they are not expected to 
propagate freely, evading thus phenomenological limits from new heavy-fermion searches. 
On the contrary, they are expected to be bound in mirror mesons which can in
principle be studied at the LHC.
 Motivation for the present study stems not only from the natural unification of all gauge couplings it provides near the Planck scale,
extending the 
spirit of \cite{PatiSalam} by including the coupling 
corresponding to the mirror generation symmetry, but also from the solution of several 
theoretical problems usually plaguing strongly-interacting BEH sectors.

First, flavour-changing neutral currents are suppressed, since the fermion-mass generation mechanism is based on a mixing 
between SM and mirror fermions, instead of
 new fermions belonging to a representation of a larger symmetry group containing 
SM fermions like in extended-technicolor models. 
This mixing mechanism, apart from generating the CKM matrix and neutrino mixing terms, 
allows for the introduction of  weak-CP violating 
phases possibly connected to the baryon asymmetry of the Universe and in parallel offers a natural solution to the strong CP problem. Second, the model does not 
create problems either with the $\Delta \rho$ parameter since the mixing operators are isospin singlets or 
with the $S$-parameter, as will become clear in Section 3. Third,  it offers a natural
see-saw mechanism explaining the smallness of neutrino masses. Forth, the
pseudo-Nambu-Goldstone bosons it predicts are not too light, since the mirror generation group self-breaks around 1 TeV and mirror-fermion chiral symmetry is broken explicitly. Last, katoptrons might provide the correct framework in order to interpret recent experimental results possibly pointing towards a strongly-interacting sector \cite{Giudice}. 
In the following, some theoretical and phenomenological aspects of katoptron theory are studied in view of current and forthcoming data from high-energy experiments.

\section{The Katoptron Lagrangian}

At energies above electroweak-symmetry breaking (around 1 TeV) and assuming a flat space-time,
 the Lagrangian ${\cal L}={\cal L}_{YM}+{\cal L}_{int}$  proposed
 is expressed  as the sum on one hand of gauge kinetic and self-interaction terms ${\cal L}_{YM}$ and 
on the other hand of  interaction terms ${\cal L}_{int}$ given by:
\begin{eqnarray}
 {\cal L}_{YM} &=& -\frac{1}{4}B_{\mu\nu}B^{\mu\nu} -\frac{1}{4}W^a_{\mu\nu}W^{a~\mu\nu}
 -\frac{1}{4}G^e_{\mu\nu}G^{e~\mu\nu} -\frac{1}{4}G^{K~e}_{\mu\nu}G^{K~e~\mu\nu} \nonumber \\
{\cal L}_{int} &=& i\sum_{j,k}\Bigg[({\bar \psi^{j|k}_u},{\bar \psi^{j|k}_d})\gamma_{\mu}{\cal D}_k^{\mu}
\left( \begin{array}{c}
  \psi^{j|k}_u \\
  \psi^{j|k}_d \end{array} \right)+
({\bar{\hat \psi}_u^{j|k}},{\bar {\hat \psi}_d^{j|k}})\gamma_{\mu}{\cal {\hat D}}_{k}^{\mu}
\left( \begin{array}{c}
{\hat \psi}_u^{j|k}\\
{\hat \psi}_d^{j|k}\end{array} \right)\Bigg]
\end{eqnarray}
\noindent where  $\gamma_{\mu}$ are Dirac matrices,
\begin{eqnarray}
B_{\mu\nu}&=&\partial_{\mu}B_{\nu}-\partial_{\nu}B_{\mu} \nonumber \\ 
W^a_{\mu\nu}&=&\partial_{\mu}W^a_{\nu}-\partial_{\nu}W^a_{\mu} -g_2f_2^{abc}W^b_{\mu}W^c_{\nu}\nonumber \\ 
G^e_{\mu\nu}&=&\partial_{\mu}G^e_{\nu}-\partial_{\nu}G^e_{\mu}-g_3f_3^{efg}G^f_{\mu}G^g_{\nu} \nonumber \\
 G^{K~e}_{\mu\nu}&=&\partial_{\mu}G^{K~e}_{\nu}-\partial_{\nu}G^{K~e}_{\mu}-g_{3K}f_3^{efg}G^{K~f}_{\mu}G^{K~g}_{\nu}
\end{eqnarray}

\noindent are the gauge-field strengths
of the symmetries $U(1)_{Y}$, $SU(2)_{L}$, $SU(3)_{C}$ and $SU(3)_{K}$ 
with coupling strengths $g_{1,2,3,3K}$ respectively, $\mu,\nu=0,...3$ 
are space-time indices,
  $f_2^{abc}$ the $SU(2)$ structure functions with $a,b,c=1,2,3$, 
 $f_3^{efg}$ the $SU(3)$, $SU(3)_{K}$ structure functions with $e,f,g=1,...8$, and
the SM-fermion generations are denoted by $j=1,2,3$.

Moreover, introducing an index $k=1,...,4$ ($k=1,2$ for SM fermions and 
$k=3,4$ for katoptrons), one defines fermion fields each consisting of two sets of 
Weyl fermions of opposite chirality:
\begin{eqnarray}
\psi^{j|k}_u &=& (N_L^j,U_L^j,N_R^{K}\delta^{3j},U_R^{K}\delta^{3j})\nonumber \\
\psi^{j|k}_d&=& (E_L^j,D_L^j,E_R^{K}\delta^{3j},D_R^{K}\delta^{3j})\nonumber \\
{\hat \psi}_u^{j|k}&=& (N_R^j,U_R^j,N_L ^{K}\delta^{3j},U_L^{K}\delta^{3j})\nonumber \\
{\hat \psi}_d^{j|k}&=&  (E_R^j,D_R^j,E_L^{K}\delta^{3j},D_L^{K}\delta^{3j})
\end{eqnarray}

\noindent where SM neutrinos, charged leptons, up-type quarks and down-type quarks are denoted by
 $N^j$, $E^j$, $U^j$ and $D^j$ respectively, the superscript ``K" denotes their mirror partners,
the 
subscripts ``L" and ``R" denote their chirality, Kronecker's $\delta^{3j}$ prevents multiple
counting of katoptron generations under summation, while the Weyl-spinor, 
color and katoptron-generation indices carried by fermions are omitted for simplicity.

 In the above, taking into account the fermion quantum numbers, 
the covariant derivatives are given by
\begin{eqnarray}
{\cal {\hat D}}_1^{\mu}&=&\partial^{\mu}+\frac{ig_1 {\hat Y}_1}{2}B^{\mu} \nonumber \\
{\cal{\hat D}}_2^{\mu}&=&\partial^{\mu}+\frac{ig_1 {\hat Y}_2}{2}B^{\mu}
+\frac{ig_3 \lambda_e}{2}G_e^{\mu} \nonumber \\
{\cal{\hat D}}_3^{\mu}&=&{\cal{\hat D}}_1^{\mu}+\frac{ig_{3K} \lambda_e}{2}G_{e}^{K~\mu} \nonumber \\
{\cal{\hat D}}_4^{\mu}&=&{\cal{\hat D}}_2^{\mu}+\frac{ig_{3K} \lambda_e}{2}G_{e}^{K~\mu} \nonumber \\
{\cal D}_k^{\mu}&=&{\cal{\hat D}}_k^{\mu}+\frac{ig_1 Y_k}{2}B^{\mu}+
\frac{ig_{2} \tau_a}{2}W_{a}^{\mu} \nonumber \\
{\rm with}~~{\hat Y}_{1}&=&\left( \begin{array}{cc}
  0 & 0\\
   0 & -1\end{array} \right), 
{\hat Y}_{2}=\left( \begin{array}{cc}
  2/3 & 0\\
   0 & -1/3 \end{array} \right), 
Y_{k}=\left( \begin{array}{cc}
  -1/2 & 0\\
   0 & 1/2\end{array} \right),
\end{eqnarray}
\noindent where the $2\times2$ unit matrix 
multiplying $\partial^{\mu}$ and the $SU(3)$, $SU(3)_K$ gauge fields is omitted, 
while $\tau_{a}$ and
 $\lambda_{e}$ are the $SU(2)$ and $SU(3)$, $SU(3)_K$ generators respectively,  omitting for 
simplicity an extra 
$U(1)^{\prime}$ interaction possibly felt only by katoptrons \cite{TriantapEJTP}. Moreover,  
neutrino Majorana mass terms responsible for a neutrino see-saw mechanism \cite{TriantapEPJ} 
are omitted, as well as possible additional
sterile-neutrino
terms implied by the embedding of this model within larger gauge symmetries \cite{TriantapEJTP}.

The Lagrangian ${\cal L}$ enjoys chiral invariance since the fermion mass-matrix $m_f$ defined later is 
initially zero. 
At energies $\Lambda_{K} \sim 1$ TeV
where $SU(3)_{K}$ becomes strongly-coupled the katoptron dynamical-mass submatrix $M$ becomes non-zero
since katoptrons acquire momentum-dependent 
dynamical constituent masses $M_i(p^2)$ similarly to QCD, which for an $SU(N_i)$
theory are 
associated to vacuum-expectation values (vevs) which may be expressed in the form
\begin{equation}
<{\bar \psi^{3|3,4}_{u,d}}{\hat \psi^{3|3,4}_{u,d}}+{\rm ~ h.c.}> ~
\approx ~ -\frac{N_i}{4\pi^2}\int dp^2 M_i(p^2)
\end{equation}
 \noindent where ``h.c." stands for ``hermitian conjugate",
 in the one-loop approximation in Landau gauge and in Euclidean space. The
 $M_i(p^2)$ are non-trivial and break chiral symmetry dynamically
only when the $SU(N_i)$ running gauge coupling exceeds a
 critical value below a certain energy due to asymptotic freedom \cite{Miransky}. 
Apart from breaking chiral symmetry, these vevs break also the electroweak gauge symmetry, providing thus the basis for a dynamical BEH mechanism. After $SU(3)_{K}$
self-breaks, the katoptron-SM fermion mixing submatrix $m$ becomes also non-zero due to gauge-invariant terms which are
forbidden at higher energies due to the unbroken katoptron generation symmetry. Diagonalization of $m_f$ gives rise to 
non-zero SM-fermion masses, to the entries of the Cabbibo-Kobayashi-Maskawa matrix and to a neutrino-mixing matrix 
\cite{TriantapEPJ},\cite{TriantapJPG}. Estimates of mirror-meson masses and of the entries of these matrices are given below, ignoring their momentum dependence.

\section{The Mirror-Meson Mass Spectrum}

The masses of the mirror mesons are not easy to estimate, large-N arguments used in \cite{EHLQ} being questionable
due to the fact that the katoptron family group $SU(3)_{K}$ becomes strongly coupled and breaks at energies
 $\Lambda_{K} \approx $ 0.5 - 1 TeV  down to $SU(2)_{K}$. Interactions corresponding to $SU(2)_{K}$ also become in
their turn strong at lower energies and break $SU(2)_{K}$. Therefore, at energies lower than $\Lambda_{K}$ where katoptrons are confined, new degrees of freedom emerge 
corresponding to two meson groups, denoted by ``A" for the lighter and "B" for the heavier
case, i.e. one expects a doubling of the
 meson mass spectrum due to the hierarchy, denoted by $r$,
 of the corresponding energy scales of mirror-fermion
chiral symmetry breaking, 
or even its tripling according to the extent by which the breaking of the
remaining $SU(2)_{K}$ family symmetry affects the corresponding mirror-meson masses.  Estimating
this hierarchy yields a factor $r$ of around
\begin{equation} 
\vspace{-2mm}
r=\exp^{}\Bigl({3 \bigl( C_2(SU(3)_{K}) - C_2(SU(2)_{K})}\bigr)\Bigr)\approx 5.75,
\end{equation}
\noindent with $C_2(g)$ the quadratic Casimir invariant of a Lie algebra $g$.
At this point, it is important to note the natural emergence of the $r$-hierarchy \cite{TriantapMPLA}
without need for any fine-tuning of parameters in the effective potential.

\subsection{The Effective Lagrangian}
The new degrees of freedom arising after dynamical symmetry breaking include 
$\pi^{K}_{A,B}=\pi^{K~l}_{A,B}t^l/2$ which are the pseudoscalar mirror meson fields corresponding to collective
operators 
\begin{equation}
{\bar {\hat \psi}^{3|3,4}_{u,d}}\gamma_5t^l\psi^{3|3,4}_{u,d}+
{\bar \psi^{3|3,4}_{u,d}}\gamma_5t^l{\hat \psi}^{3|3,4}_{u,d}
\end{equation}
\noindent with $t^{l}$, $l=1,...64$
 the generators of the broken $SU(8)$ axial chiral symmetry discussed later plus an index corresponding to the $U(1)_A$ axial symmetry and the
$\eta^{\prime}$ meson in QCD, with $M_{\pi^K_{A,B}}$ being their mass matrix and
$g^{j|k=1,2}_{A,B|u,d}$ their effective couplings to SM fermions. In addition, one has
 Higgs-type scalar fields
$\sigma^K_{A,B}$ of mass $M_{\sigma^K_{A,B}}$
corresponding to collective operators
 \begin{equation}
{\bar {\hat \psi}^{3|3,4}_{u,d}}\psi^{3|3,4}_{u,d}+
{\bar \psi^{3|3,4}_{u,d}}{\hat \psi}^{3|3,4}_{u,d}
\end{equation}
analogous to fields sometimes referred to as ``techni-dilatons" in 
the technicolor literature \cite{Yama}, although in our case the lightness of $\sigma^K_A$
has a different origin.
 
Moreover, one has vector fields
$\rho^K_{\mu~A,B}= \rho^{K~l}_{\mu~A,B}t^l/2$
  corresponding to collective operators
 \begin{equation}
{\bar \psi^{3|3,4}_{u,d}}\gamma_{\mu}t^l\psi^{3|3,4}_{u,d}+
{\bar {\hat \psi}^{3|3,4}_{u,d}}\gamma_{\mu}t^l{\hat \psi}^{3|3,4}_{u,d}
\end{equation}
\noindent  and axial-vector fields
$a^K_{\mu~A,B}= a^{K~l}_{\mu~A,B}t^l/2$
 corresponding to collective operators
 \begin{equation}
{\bar \psi^{3|3,4}_{u,d}}\gamma_{\mu}\gamma_5t^l\psi^{3|3,4}_{u,d}+
{\bar {\hat \psi}^{3|3,4}_{u,d}}\gamma_{\mu}\gamma_5t^l{\hat \psi}^{3|3,4}_{u,d}.
\end{equation}

\noindent One may also define subgroups $\rho^{K}_{\mu~k=1,2}$, $a^K_{\mu~k=1,2}$
of these operators, i.e.
$\rho^{K}_{\mu~1}$, $a^K_{\mu~1}$ including singlets and color-singlet isospin triplets, and
$\rho^{K}_{\mu~2}$  including, in addition to $\rho^{K}_{\mu~1}$, neutral isospin-singlet
vector color-octets. 
All  possible
combinations of meson quantum numbers will be listed later.

Some of the interactions of the new degrees of freedom
 can 
be studied in principle by a lowest-order effective chiral Lagrangian ${\cal L}_{eff}$, after integrating out 
the katoptrons and the katoptron-generation gauge fields, which, omitting amongst others 
the masses and field strengths
of the various vector and axial-vector fields, is given by
\begin{eqnarray}
{\cal L}_{eff}&=&\sum_{n}\Bigg[
\Sigma^2_n\Bigg(\frac{1}{2}({\cal D}^{\mu}\sigma^K_{n})^2 +
\frac{F^2_{n}}{4}{\rm tr~}\Bigg\{({\cal D}^{\mu}\Pi_{n})^{\dagger}
 {\cal D}_{\mu}\Pi_{n}+
M^2_{\pi^K_n}\Pi_{n}^{\dagger}+{\rm h.c.}\Bigg\}\Bigg)
+\nonumber \\&&+
\frac{F^2_{n}M^2_{\sigma^K_n}}{8}\left(1-(\Sigma^2_{n}-1)^2\right)
+
\frac{F_n}{2}\sum_{j,k}{\rm tr~}
\Bigg\{(g^{j|k}_{n|u}{\bar{\hat \psi}_u^{j|k}},
g^{j|k}_{n|d}{\bar {\hat \psi}_d^{j|k}})\left(\Sigma_{n}+
\Pi_n\right)\left( \begin{array}{c}
  \psi^{j|k}_u \\
  \psi^{j|k}_d \end{array} \right)+{\rm h.c.}\Bigg\}\Bigg]+\nonumber \\
&&+{\rm tr~}\Bigg\{F_B g_{BA}(\Sigma_{B}+\epsilon\Pi_B)
\Bigg(\frac{1}{2}({\cal D}^{\mu}\sigma^K_{A})^2+
\frac{F^2_A}{4}({\cal D}^{\mu}\Pi_{A})^{\dagger}{\cal D}_{\mu}\Pi_{A}
\Bigg)+{\rm h.c.}\Bigg\} \nonumber \\
&&+i\sum_{j,k}
\Bigg[({\bar \psi^{j|k}_u},{\bar \psi^{j|k}_d})\gamma_{\mu}{\tilde {\cal D}}_k^{\mu}
\left( \begin{array}{c}
  \psi^{j|k}_u \\
  \psi^{j|k}_d \end{array} \right)+
({\bar{\hat \psi}_u^{j|k}},{\bar {\hat \psi}_d^{j|k}})\gamma_{\mu}
 {\cal {\tilde{\hat D}}}_{k}^{\mu}
\left( \begin{array}{c}
{\hat \psi}_u^{j|k}\\
{\hat \psi}_d^{j|k}\end{array} \right)\Bigg]
\end{eqnarray}
\noindent  where the summation runs over the two mirror-meson groups $n=A,B$, the SM-fermion generation index $j=1,2,3$ and
the SM lepton and quark index $k=1,2$,  $\epsilon$ is a CP-violation
parameter matrix related to the mass-generation mechanism and 
 $F_{A,B}$ are mirror-meson decay
constants.
The characteristic scales $F_{A,B}$ introduced here
 are analogous to the pion decay constants $f_{\pi}$ in QCD,
 they are each assumed to be shared by all the members of 
the respective mirror-meson groups for simplicity,
and they are studied later.
 Moreover, a simple ansatz for the
symmetry-breaking potentials has been employed. 
Furthermore, it is assumed in the following
that the strength of the interaction of group-``B" with group``A" mesons is
measured by 
\begin{equation}
g_{BA} \sim F_{A}/F_B \sim 1/r.
\end{equation}

The following definitions have been made in the expression for the effective Lagrangian above:
\begin{eqnarray}
\Sigma_{A,B}&\equiv& 
\exp{(\sigma^{K}_{A,B}/F_{A,B})}, \hspace{3.6cm} \Pi_{A,B}\equiv
\exp{(2i\pi^{K}_{A,B}/F_{A,B})} \nonumber \\
{\cal D}^{\mu}\Pi_{A,B}&=&
(\partial^{\mu}-i{\tilde R}^{\mu}) \Pi_{A,B}+i \Pi_{A,B}{\tilde L}^{\mu} \nonumber \\
{\cal {\tilde {\hat D}}}_k^{\mu} &=&\partial^{\mu}+ {\tilde R}^{\mu}_k, \hspace{4.9cm} 
{\tilde {\cal D}}_k^{\mu} =\partial^{\mu}+ {\tilde L}_k^{\mu}\nonumber \\
{\tilde R}_k^{\mu}&=&R_k^{\mu}+\rho_k^{K~\mu}+a_1^{K~\mu}, 
\hspace{3.5cm} {\tilde L}_k^{\mu} = 
L_k^{\mu}+\rho_k^{K~\mu}-a_1^{K~\mu} \nonumber \\
R_1^{\mu} &=&
ieQ_1 r^{\mu}, ~~~~~~R_2^{\mu}=
ieQ_2 r^{\mu} +
\frac{ig_3\lambda_e}{2}G^{\mu}_e, \hspace{0.4cm}  L_k^{\mu}=
 R_k^{\mu}-\frac{ie}{2}X^{\mu}_{k} \nonumber \\
Q_{1,2}&=&{\hat Y}_{1,2}, \hspace{1.2cm}r^{\mu}= A^{\mu}-\tan{\theta_w}Z^{\mu},
\hspace{1.2cm}e\equiv g_2\sin{\theta_w} \nonumber \\
X^{\mu}_k&=&\frac{1}{\sin{\theta_w}}\left( \begin{array}{cc}
 - Z^{\mu}/\cos{\theta_w} & W^{+~\mu}{\cal M}_k\\
   W^{-~\mu}{\cal M}_k^{\dagger} & Z^{\mu}/\cos{\theta_w}\end{array} \right)\nonumber \\
\left( \begin{array}{c}
  A^{\mu} \\
   Z^{\mu} \end{array} \right)&=&\left( \begin{array}{cc}
  \sin{\theta_w} & \cos{\theta_w}\\
 \cos{\theta_w}  & -\sin{\theta_w}\end{array} \right)
\left( \begin{array}{c}
  W_3^{\mu} \\
   B^{\mu} \end{array} \right), \hspace{1.6cm}W^{\pm~\mu}=W_1^{\mu}\mp iW_2^{\mu} 
 \end{eqnarray}

\noindent 
where $\theta_w$ is the Weinberg angle,
 $A^{\mu}$, $e$ and $Q_{1,2}$ are the usual electromagnetic field, coupling and charges,
$Z^{\mu}$ and $W^{\pm~\mu}$ the usual massive fields corresponding to the broken
electroweak gauge symmetry, ${\tilde R^{\mu}}$ and  ${\tilde L^{\mu}}$ are
natural embeddings of ${\tilde R_k^{\mu}}$ and 
 ${\tilde L_k^{\mu}}$ within  $SU(8)$, while
${\cal M}_1$ and ${\cal M}_2$ are  the neutrino and CKM mixing matrices respectively.

The effective Lagrangian
above should contain all the necessary information needed to describe the mirror 
mesonic spectrum and its interactions at lowest order, 
taking care to work with each of the sectors $A,B$ at their particular range of valid energies characterized by $F_{A,B}$, since it is non-renormalizable.
For this reason, the expression
  arising from an interchange of the $A,B$ subscripts of the term multiplied by $g_{BA}$ 
is omitted, assuming that group-``B" mirror mesons decouple at lower energies on the order of group-``A" meson masses
where the field $\sigma^K_{A}$ may be studied. This point will be further discussed in Section 4.
In any case, the self-breaking of $SU(3)_{K}$ violates chiral symmetry to an extent that might invalidate the chiral expansion and restrain the applicability of this method. 

Next comes the definition of
 the mass matrix $m_{f}$ pertaining to both SM fermions and katoptrons, giving rise to the
couplings $g^{j|k}_{n|d}$ in the effective Lagrangian 
and describing roughly the basis for the fermion mass-generation 
mechanism explained in detail elsewhere
\cite{TriantapEPJ}, \cite{TriantapJPG}. Before chiral symmetry
breaking, $m_f$ is obviously trivial. When $M_{A,B} \neq 0$ however, we define:
\begin{equation}
m_f=\left( \begin{array}{cc}
m_{SM} & m  \\
m & M  \end{array} \right){\rm ~,with~}
M\equiv\left( \begin{array}{cc}
M_A & 0  \\
0 & M_B  \end{array} \right)~{\rm ~and~}
m\equiv\left( \begin{array}{cc}
m_{AA}  & m_{AB}  \\
m_{AB} & m_{BB}  \end{array} \right)
\end{equation}
\noindent being
the katoptron dynamical-mass matrix $M$ and the SM-fermion $\&$ katoptron mixing matrix $m$
respectively, where it is assumed that
\begin{equation}
m_{AB} \sim m_{BB}/r. 
\end{equation}

Before diagonalization, one has $m_{SM}=0$ and $m_{AA}=0$. Diagonalization of $m_f$,
apart from giving rise to the mixing matrices ${\cal M}_{1,2}$ defined previously,
 yields approximately
\begin{equation}
m^D_{SM} =
\left( \begin{array}{cc}
m_{f_{A}} & 0  \\
0 & m_{f_{B}}  \end{array} \right) 
\approx \left( \begin{array}{cc}
(m^{D}_{AA})^{2}/M_A & 0  \\
0 & m^2_{BB}/M_B  \end{array} \right) 
\end{equation}
\begin{equation}
{\rm where}~~~m^{D}_{AA} \approx m_{BB}/r^2, 
\end{equation}
\noindent resulting finally in the SM-fermion intra-generation mass hierarchy
\begin{eqnarray}
m_{f_B} &=& m^2_{BB}/M_B \nonumber \\
m_{f_A} &=&  m_{f_B}/r^3.
\end{eqnarray}

\noindent Taking $M_B = 1$ TeV and $m_{BB} = 0.418$ TeV gives $m_{f_B} \approx 0.175 {\rm ~ TeV} \approx m_t$, and 
$m_{f_A} \approx 0.92$ GeV $\approx m_c$, which are correct orders of magnitude for 
the SM-fermion masses of the two heavier generations.
 A similar mechanism is at works regarding the lightest fermion generation. 
One may introduce CP-violating phases in this matrix which
are linked to the $\epsilon$ matrix introduced in the effective Lagrangian above and 
might provide the necessary mechanism explaining the baryon asymmetry of the 
Universe.

\subsection{Meson Mass Estimates}
A more detailed discussion of mirror mesons follows next.
Similarly to SM fermions, each generation of katoptrons consists of $N=8$ fermions or $N_D = N/2 = 4$
 isospin (SU(2)) doublets, 
i.e. one lepton (color-singlet) 
doublet and one quark (color-triplet) doublet. This gives rise to a  
chiral symmetry of the initial Lagrangian described by SU(N)$_L \otimes$SU(N)$_R$ with $N=8$. In terms of its
 fundamental representation, the adjoint representation can be decomposed under SU(2)$\times$SU(3) in the 
following way:
\begin{eqnarray}
\bf{[8_{L}]\otimes [8_{R}]} &=& \bf{[2 \times (3+1)]\otimes [2\times (\bar{3} + 1)]} \nonumber \\ 
&=& \bf{(3_2+1)\times (3\otimes \bar{3} + 3 + \bar{3} + 1) =}
\bf{(3_2 + 1)\times (8 + 1 + 3 + \bar{3} +1)}
\end{eqnarray}

\noindent with $\bf{2\otimes 2 = 3_2 + 1}$ for weak SU(2) ($\bf{3_2}$ denoting an isospin triplet) 
and $\bf{3\otimes \bar{3} = 8 + 1}$ for color SU(3).

Strong dynamics of the katoptron generation group around 1 TeV 
lead to the breaking of the chiral symmetry down to its diagonal vector subgroup, followed by the formation of 
$N^2-1$ Nambu-Goldstone (NG) bosons corresponding to the broken axial-vector symmetry. 
Apart from these
pseudoscalar particles which are analogous to the lightest QCD mesons, one also expects spin-1 resonances
analogous to the $\rho$ meson. The formula above provides a simple counting device of these mesons according to 
their quantum numbers. The following analysis bears resemblance to technicolor models \cite{Fahri}, and 
one can list the expected mirror mesons following the decomposition above, 
with the superscript ``K"
a reminder of their mirror-fermion content.

First, one has three  
 mirror pions, $\pi^{K~b~0}$ and $\pi^{K~b~\pm}$ 
``eaten" by the electro-weak gauge bosons, becoming  thus the longitudinal components of $W^{\pm}$
and $Z^{0}$. Then, one has five more (for a total of eight) color-singlets, including
$\eta^{\prime K}$ corresponding to a broken $U(1)_A$ symmetry:
\begin{equation}
\pi^{K~a~0}, ~\pi^{K~a~\pm}, ~\pi^{K~0\prime},  \eta^{\prime K} {\rm ~~(spin-0)~~}
{\rm ~and~~}
\rho^{K~a,b~0}, ~\rho^{K~a,b~\pm}, ~\rho^{K~0\prime} {\rm ~and~~} \omega^{K} {\rm~~~(spin-1)},
\end{equation}
\noindent four color-triplets (usually called ``leptoquarks") together with their anti-particles,
\begin{equation}
\pi^{K~1,2,2\prime,5}_3, ~{\bar \pi^{K~1,2,2\prime,5}_3} {\rm ~(spin-0)~~~and~~~}
\rho^{K~1,2,2\prime,5}_{3}, ~{\bar \rho^{K~1,2,2\prime,5}_{3}} {\rm ~(spin-1)}, 
\end{equation}

\noindent all of them fractionally charged 
(either $-\frac{1}{3}$ for $\pi_3^{K~1}$, $\rho^{K~1}_3$, or $\frac{2}{3}$
for $\pi_3^{K~2,2\prime}$, $\rho^{K~2,2\prime}_3$, or $\frac{5}{3}$ for $\pi_3^{K~5}$, $\rho^{K~5}_3$),
and four are color-octets,  denoted by 
\begin{equation}
\pi^{K~0}_8, ~\pi^{K~\pm}_8, ~\pi^{K~0\prime}_8  {\rm ~~~~~(spin-0)~~~and~~~} \rho^{K~0}_8, ~\rho^{K~\pm}_8, 
~\rho^{K~0\prime}_8  {\rm ~~~~~(spin-1)}.
\end{equation}

\noindent Note that isospin-singlet mesons apart from $\omega^{K}$ are denoted  
 by primed symbols above, 
the rest being members of isospin triplets. Another obvious fact to bear in mind is
 that mirror mesons with equal charge and color may mix with each other, 
like $\pi^{K~a~0}$ with $\pi^{K~0\prime}$, 
$\pi_3^{K~2}$ with $\pi_3^{K~2\prime}$ and
  $\pi_8^{K~0}$ with $\pi_8^{K~0\prime}$, and similarly for their vector-meson counterparts. These are the
most obvious lightest mirror mesons one expects, without excluding the existence of their parity partners (scalars and
axial vectors) and heavier excited states of all these combinations.

 In particular, one should not forget the mirror
analogue of the $\sigma$ scalar QCD resonance, i.e. $\sigma^K$,
 the lowest-lying resonance heavier than the three pseudo-scalar pions, the analogues of which
are here ``eaten" by $W^{\pm}$ and $Z^{0}$.
Having the same quantum numbers, $\sigma^K$ corresponds to the
``Higgs-type" particle recently discovered at the LHC.
The fact that the mass of the scalar  particle detected is lower than double the masses not only of the 
electro-weak gauge bosons but of the top-quark as well
might partially explain its relatively small width compared to the one of the sigma meson in QCD, which mainly
decays into two pions.
Lest $\pi^{K~a~0}, ~\pi^{K~a~\pm}$ are finally not observed, one might 
 consider the possibility of
their mixing with $\pi^{K~b~0}, ~\pi^{K~b~\pm}$, which would mean that they are also "eaten" by the $Z^0,W^{\pm}$
gauge bosons. In principle, composite states consisting of more
than two katoptrons (like mirror protons and mirror neutrons) 
are also possible, even though they should be harder to produce at particle colliders. 

However, as has been noted already, 
this spectrum is doubled or even tripled due to the breaking of the katoptron-generation 
symmetry. In the following, the lighter mirror mesons, denoted  by $\pi^{K}_A$ and $\rho^{K}_A$,
omitting numerical superscripts and color subscripts,  
correspond to katoptrons of the two lighter
mirror generations expected to have dynamical masses of around 
\begin{equation}
\Lambda_K/r \approx M_A \approx 0.1 - 0.2 {\rm ~TeV}, 
\end{equation}
\noindent bearing in mind that they might be further
split into two distinct subgroups according to the masses of their mirror-fermion content. Similarly, 
 the heavier mirror mesons are denoted by $\pi^{K}_B$ and $\rho^{K}_B$, 
corresponding to katoptrons with constituent masses 
\begin{equation}
M_B =rM_A\sim 0.57 - 1.15 {\rm ~TeV}.
\end{equation}

\noindent The range of these
masses is constrained via the Pagels-Stokar formula \cite{PS} which should reproduce the correct
order of magnitude for the weak scale 
$v \approx 246$ GeV  with three generations of $N_D=4$
doublets having masses $M_i$, $i=A,B$, for a strongly-coupled $SU(N_i)$
theory with a $\Lambda_{K~i}$ momentum cut-off:
\begin{equation}
v \approx \frac{1}{\pi}\sqrt{\sum_{i=A,B}N_{i}M^2_i  \ln{(\Lambda^2_{K~i}/M^2_i)}},
\end{equation}

\noindent where $N_{A}=2\cdot2=4$ and $N_{B}=3$. 
 The fact that $M_B$ is much larger than $M_A$ implies that the value of the weak scale is mainly determined 
by third-generation katoptrons. 

Since the katoptron generation group is broken, one has
 to introduce different decay constants for the mirror pions according to their mass, which are
denoted by $F_{A}$ and $F_{B}$. Due to the fact that the decay constants
are roughly proportional to the katoptron masses they correspond to (up to logarithmic corrections), and that $M_{B}$
 is quite larger than $M_{A}$, one expects that 
\begin{eqnarray}
M_B /\pi \approx F_{B} &\approx& \frac{v}{2\sqrt{1+2/r^2}} \approx 120 {\rm~ GeV} \nonumber \\
M_A /\pi \approx F_{A} &\approx& F_{B}/r \approx 21 {\rm~ GeV}.
\end{eqnarray}

\noindent As will soon become clear however, the katoptron effective couplings are chosen
here  in a way that final expressions 
for the various cross-sections depend on $v$ instead of $F_{A,B}$.

Furthermore, order-of-magnitude estimates based on QCD 
for spin-1 mirror mesons give masses of around 
\begin{equation}
M_{\rho^K_{A,B}} \approx \frac{m_{\rho}}{m_{u}}M_{A,B}, 
\end{equation}

\noindent with $m_{\rho} \approx$ 770 MeV the mass of the $\rho$ meson and $m_{u} \approx$ 313 MeV
 the constituent mass of the up quark. Therefore, one may estimate
\begin{eqnarray}
M_{\rho^K_{A}} &\approx& 0.25 - 0.5 {\rm~ TeV} \nonumber \\
M_{\rho^K_{B}} &\approx&  1.4 - 2.8 {\rm~ TeV}, 
\end{eqnarray}

\noindent and it is assumed that 
$\eta^{\prime K}_{A,B}$ and $\omega^{K}_{A,B}$ fall within the same respective ranges.
Relevant Tevatron exclusion limits for vector-resonance masses below about 500 GeV might be circumvented by
non-QCD-like dynamics \cite{Zerwekh}, increasing the relevant masses or decreasing the relevant effective
couplings, which is not unreasonable taking into account that the katoptron generation group
breaks after it becomes strongly coupled.

A short note regarding the $S$ parameter \cite{Peskin}
 is in order, in view of the large number of new chiral fermions introduced in the theory. Two major 
contributions to this parameter are generally expected 
due to group-``A" and group-``B" spin-1 resonances, i.e. 
\begin{equation}
S=S_A+S_B=4\pi\sum_A\Bigg(\frac{F^{2}_{\rho_A^{K}}}{M^{2}_{\rho_A^{K}}}-
\frac{F^{2}_{a_A^{K}}}{M^{2}_{a_A^{K}}}\Bigg)+
4\pi
\sum_B\Bigg(\frac{F^{2}_{\rho_B^{K}}}{M^{2}_{\rho_B^{K}}}-\frac{F^{2}_{a_B^{K}}}{M^{2}_{a_B^{K}}}\Bigg).
\end{equation} 
\noindent Assuming that  the first Weinberg sum rule (WSR) is dominated by group-``B" mesons,
which is reasonable since $F_{B} \sim 6F_A$, one finds
\begin{equation}
\sum_B F^2_{\rho^K_B}-F^2_{a^K_B}  \approx v^2.
\end{equation}
Moreover, assuming that vector and axial-vector
meson masses are approximately equal, i.e. $M_{\rho^K_B}\approx M_{a^K_B}$, one has very roughly
\begin{equation}
S_B \approx 4\pi (v/M_{\rho^K_B})^2~^{<}_{\sim~} 0.122
\end{equation}

\noindent for $M_{\rho^K_{B}}>2.5$ TeV. On the other hand, since the decay constants $F^{2}_{\rho^K_A, a_A^{K}}$
of group-``A" spin-1 resonances
is not severely constrained by WSR, $S_A$ can be very close to zero or even negative. This scheme has therefore the
potential to offer an elegant solution to the $S$-parameter problem providing non-QCD-like dynamics 
without an unnatural fine-tuning of various
parameters and obviating usual stringent limits on the number of new chiral fermions.

The masses of spin-0 mirror mesons are studied next. 
Pseudoscalar mesons should be relatively light if considered as Nambu-Goldstone 
bosons of the broken mirror chiral symmetry. 
However, since the gauged mirror family symmetry is consecutively broken, these are massive
pseudo-NG bosons. It is assumed in the following that  mirror-pion
and $\sigma^{K}$ masses 
are grouped either around
\begin{equation}
100 - 200 {\rm ~GeV} ~~~{\rm for~}(\sigma^{K}_A, \pi^{K}_A) 
\end{equation}
\noindent or 
around 
\begin{equation}
0.57 - 1.15 {\rm ~ TeV} ~~~{\rm for~} (\sigma^{K}_B, \pi^{K}_B).
\end{equation} 
\noindent The assumed proximity of the mirror-pion and mirror-sigma
masses implies that the explicit 
breaking of the chiral symmetry here is relatively more significant than the corresponding one in QCD.
Note that colored mirror pions receive additional contributions to their masses due to QCD, which
are on the order of 
\begin{equation}
\sqrt{\alpha_s}M_{\rho^K_{A,B}} {~~~\rm~for~}\pi^{K}_{8~A,B}
\end{equation} and
\begin{equation}
\frac{2}{3}\sqrt{\alpha_s}M_{\rho^K_{A,B}}{~~~\rm~for~}\pi^{K}_{3~A,B}, 
\end{equation}
where $\alpha_s$ is the
value of the QCD coupling near the mirror-pion mass \cite{Preskill}. 

The considerations above lead to the following estimates:
\begin{eqnarray}
M_{\pi^{K}_{3~A,B}} &\approx& \sqrt{M^2_{\pi^{K}_{A,B}}+\frac{4}{9}\alpha_sM^2_{\rho^K_{A,B}}}  \nonumber \\
M_{\pi^{K}_{8~A,B}} &\approx& \sqrt{M^2_{\pi^{K}_{A,B}}+\alpha_sM^2_{\rho^K_{A,B}}}
\end{eqnarray}
\begin{figure}
\vspace{-3cm}
\hspace{-3cm}
\centering
\includegraphics
[angle=0,width=19cm]
{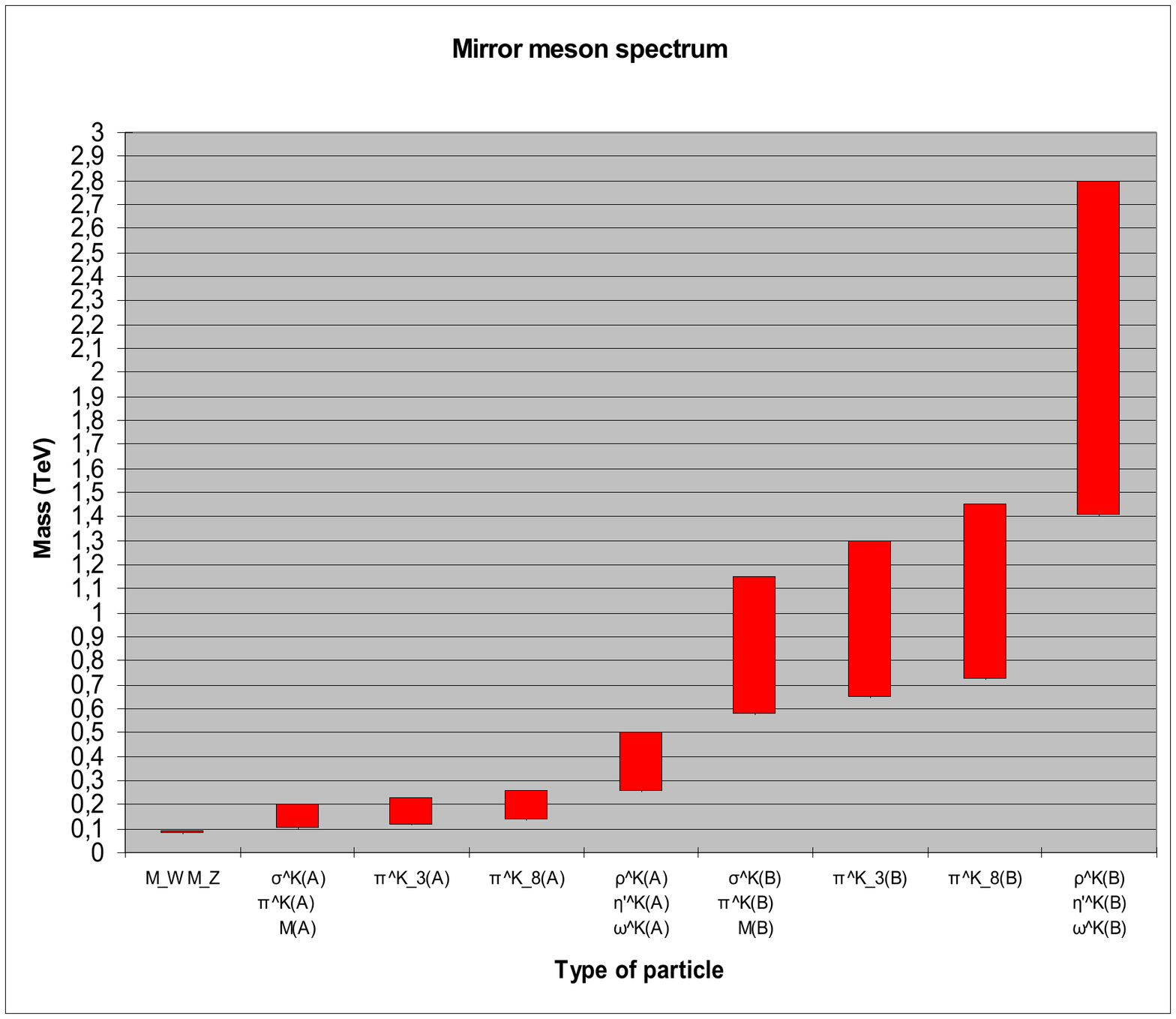}
\vspace{-10cm}
\caption{Rough order-of-magnitude estimates of mirror-meson mass expected 
ranges, where \hfill \break 
$M_{-}W,Z: M_{W,Z}$ \hfill \break 
$\sigma^{\wedge}K(A,B): \sigma^{K}_{A,B}$ \hfill \break
$\pi^{\wedge}K(A): \pi^{K~a~0}_{A}, ~\pi^{K~a~\pm}_{A}, ~\pi^{K~0\prime}_{A}$ \hfill \break
$\pi^{\wedge}K(B): \pi^{K~a,b~0}_{B}, ~\pi^{K~a,b~\pm}_{B}, ~\pi^{K~0\prime}_{B}$ \hfill
\break 
$M(A,B): M_{A,B}$ \hfill \break
$\pi^{\wedge}{K}_{-}3(A,B):\pi^{K~1,2,2\prime,5}_{3~A,B}, ~{\bar \pi^{K~1,2,2\prime,5}}_{3~A,B}$ \hfill
\break $\pi^{\wedge}{K}_{-}8(A,B): \pi^{K~0}_{8~A,B}, ~\pi^{K~\pm}_{8~A,B}, ~\pi^{K~0\prime}_{8~A,B}$ 
and \hfill \break
$\rho^{\wedge}K(A,B)$: all kinds of $\rho^{K}_{A,B}$ mesons \hfill \break
$\eta^{\prime \wedge}K(A,B): \eta^{\prime K}_{A,B}$ \hfill \break
$\omega^{\wedge}K(A,B): \omega^{K}_{A,B}$}
\end{figure}
\noindent Therefore, one expects colored mirror meson masses to be approximately given by 
\begin{eqnarray}
M_{\pi^K_{3~A}} &\sim&  0.11 - 0.23 {\rm~TeV}\nonumber \\
M_{\pi^K_{3~B}} &\sim&  0.64 - 1.3 {\rm~TeV}\nonumber \\
M_{\pi^K_{8~A}} &\sim&  0.13 - 0.26 {\rm~TeV} \nonumber \\
M_{\pi^K_{8~B}} &\sim&  0.72 - 1.45 {\rm~TeV}
\end{eqnarray}

 These rough order-of-magnitude estimates can be visualized in Fig.1 which is mainly indicative, keeping
in mind that experiments might reveal non-negligible deviations from these values should these mesons exist.
For instance, the $\eta^{K~\prime}_{A,B}$  mesons might have masses closer to the ones of color-singlet mirror
pions, around half of what is indicated in Figure 1.
Having described the results of the dynamical mass-generation mechanism, 
we study below decays of the new degrees of freedom arising at energies 
testable at the LHC which have particularly
interesting phenomenological implications.

\section{Mirror-Meson Processes}

\subsection{Decay Widths}
The breaking of the strong mirror-generation group renders all mirror mesons, together with the 
katoptrons they consist of, unstable. Therefore, apart from inferring the existence of katoptrons from quantum corrections 
to various processes, direct detection of mirror mesons via their decays is crucial for the falsifiability of the theory. 
It is well known that the cross-section $\sigma$
of a proton-antiproton collision ${\bar p} p$ with center-of-mass energy $\sqrt{s}$ resulting in a final state $X$
via a resonance $R$ for a specific invariant-mass bin defined by 
$(\tau_{min},\tau_{max})$ is given by
\begin{equation}
\sigma({\bar p} p \longrightarrow R  \longrightarrow X) = \int^{\tau_{max}}_{\tau_{min}}
 d \tau \sum_{\alpha, \beta}\frac{dL_{\alpha \beta}}{d\tau}{\hat \sigma(\alpha \beta  \longrightarrow R \longrightarrow X)}
\end{equation}

\noindent where $\tau\equiv{\hat s}/s$ is the product of the two proton-energy fractions carried by partons $\alpha$ and $\beta$, 
$\sqrt s = 13$ TeV is the LHC RUN II center-of-mass energy which is studied in this work and 
\begin{eqnarray}
\frac{dL_{\alpha \beta}}{d \tau} &\equiv& \int_{\tau}^1 \frac{dx}{x}P_{\alpha}(x)P_{\beta}(\tau/x) \nonumber \\
{\hat \sigma(\alpha \beta  \longrightarrow R \longrightarrow X)} &\equiv&
\frac{4\pi c }{c_{\alpha}c_{\beta}}
\frac{\Gamma^{\alpha \beta}\Gamma^X}{({\hat s}-M^2)^2+M^2\Gamma^2_{tot}}
\end{eqnarray}

\noindent 
where $P_{\alpha, \beta}$ are the parton distribution functions, $M$ and $\Gamma_{tot}$ are the mass and the total 
width of the resonance $R$, and $\Gamma^{\alpha \beta}, \Gamma^X$ are the production and decay widths of $R$.  Moreover, ${\hat \sigma}$ is given by 
the relativistic Breit-Wigner formula, with $c, c_{\alpha, \beta}$  appropriate color factors. 

The large collision energy at the LHC, in conjunction with the fact that the
production cross-section
 of mirror mesons from up quarks ${\hat \sigma}_{{\bar u}u}$ for instance
 is several orders of magnitude smaller than 
their production cross-section from gluons ${\hat \sigma}_{gg}$ {\it ceteris paribus}, i.e.
\begin{equation}
\frac{{\hat \sigma}_{{\bar u}u}}{{\hat \sigma}_{gg}}~^{<}_{\sim}~
9\left(\frac{\pi m_u}{\alpha_s M_{\pi^K}}\right)^2 \approx 2\times 10^{-5}
\end{equation}
 renders the use of just the gluon distribution functions a fairly good approximation. Approximating the CTEQ6L1 gluon-luminosity data for $\sqrt{s}=13$ TeV \cite{Quigg} 
by a fitting function for $\frac{dL_{gg}}{d\tau}$ and 
using the narrow-width approximation for ${\hat \sigma(gg \longrightarrow X)}$ yields an order-of-magnitude estimate
of the total cross-section: 
\begin{eqnarray}
\sigma({\bar p} p  \longrightarrow R \longrightarrow X) &=& {\cal L}(M)
 \frac{c\Gamma^{gg}\Gamma^X}{M\Gamma_{tot}}\nonumber \\
 {\rm with}~~{\cal L}(M) &\approx& 4(M/{\rm TeV})^{-(6+1.6\log_{10}(M/{\rm TeV}))} {~~\rm nb},
\end{eqnarray}

\noindent where ${\cal L}$ is slightly underestimated to account for various experimental inefficiencies. An effort is made next to identify the most important mirror-meson decay widths,
before reporting the relevant cross-section estimates.
Since the analysis that follows concentrates on the gluon-fusion production mechanism, 
interesting processes depending on quark distribution functions
are left for future investigations, like the production of color-singlet
spin-1 mirror-mesons decaying to electroweak gauge bosons:
\begin{equation}
\vspace{-2mm}
{\bar q_i}q_j \longrightarrow \rho^{K~b~\pm}_{A, B} \longrightarrow W^{\pm}+Z^{0},
\end{equation}

\noindent which might potentially explain relevant excesses detected recently \cite{excess}. 

It is assumed that the decay widths of generic mirror mesons to
SM fermions are given by the following Higgs-like expressions: 
\begin{eqnarray}
\Gamma(\sigma^K_A, \pi^K_A \longrightarrow {\bar f_A}f_A) 
&\approx& \frac{c_f}{8\pi^3} \frac{{\tilde m}^4_{AA}}{ F^2_A v^2} M_{\sigma_{A}^K, \pi_{A}^K}
\approx \frac{c_f m^2_{{f_A}}M_{\sigma_{A}^K, \pi_{A}^K}}{8\pi v^2}\nonumber \\
\Gamma(\sigma^K_B, \pi^K_B \longrightarrow {\bar f_B}f_B) 
&\approx& \frac{c_f}{8\pi^3}\frac{ m^4_{BB}}{ F^2_B v^2} M_{\sigma_{B}^K, \pi_{B}^K}
\approx \frac{c_f m^2_{{f_B}}M_{\sigma_{B}^K, \pi_{B}^K}}{8\pi v^2}\nonumber \\
\Gamma(\sigma^K_A, \pi^K_A \longrightarrow {\bar f_B}f_B) 
&\approx& \frac{c_f}{8\pi^3}\frac{ m^4_{AB}}{ F^2_A v^2}\frac{M_B^2}{M_A^2} M_{\sigma_{A}^K, \pi_{A}^K}
\approx \frac{c_f m^2_{{f_B}}M_{\sigma_{A}^K, \pi_{A}^K}}{8\pi v^2} \nonumber \\
\Gamma(\sigma^K_B, \pi^K_B \longrightarrow {\bar f_A}f_A) 
&\approx& \frac{c_f}{8\pi^3}\frac{m^4_{AB}}{ F^2_B v^2}\frac{M_A^2}{M_B^2} M_{\sigma_{B}^K, \pi_{B}^K}
\approx \frac{c_f m^2_{{f_A}}M_{\sigma_{B}^K, \pi_{B}^K}}{8\pi v^2}
\end{eqnarray}

\noindent where $c_f=3$ when the meson is a color singlet and the final fermions are quarks, with $c_f=1$ otherwise.

To list specific examples, one might start with neutral spin-0 mirror mesons decaying into pairs of third-generation SM fermions which 
are more interesting due to their heaviness:
\begin{eqnarray}
\Gamma (\sigma_{A}^{K}, \pi^{K~0\prime}_A \longrightarrow {\bar b} b) &\approx&
\frac{3 m^2_b M_{\sigma_A^{K}, \pi_A^{K~0\prime}}}{8\pi v^2} = 3.5-7 {\rm ~MeV}
\equiv \Gamma_{tot~(\sigma_{A}^{K}, \pi^{K~0\prime}_A)}/2\nonumber \\
\Gamma (\pi_{8~A}^{K~0\prime} \longrightarrow {\bar b} b) &\approx&
\frac{m^2_b M_{\pi^{K~0\prime}_{8~A}}}{8\pi v^2} = 2.3-4.6{\rm ~MeV}\equiv \Gamma_{tot~(\pi^{K~0\prime}_{8~A})}/2 \nonumber \\
\Gamma (\sigma_{A,B}^{K}, \pi^{K~0\prime}_{A,B} \longrightarrow {\bar \tau} \tau) &\approx&
\frac{m^2_{\tau}M_{\sigma_{A,B}^{K}, \pi_{A,B}^{K~0\prime}}}{8\pi v^2}
= (0.2-0.4~ (A),1.2-2.4~ (B)){\rm ~MeV} \nonumber \\
\Gamma (\sigma_{B}^{K}, \pi^{K~0\prime}_B \longrightarrow {\bar t} t) &\approx&
\frac{3 m^2_t M_{\sigma_B^{K}, \pi^{K~0\prime}_B}}{8\pi v^2}= 34-68 {\rm ~GeV} \equiv \Gamma_{tot~(\sigma_{B}^{K}, \pi^{K~0\prime}_B)}/2 \nonumber \\
\Gamma (\pi_{8~B}^{K~0\prime} \longrightarrow {\bar t} t) &\approx&
\frac{m^2_t M_{\pi^{K~0\prime}_{8~B}}}{8\pi v^2} = 20-40 {\rm ~GeV}\equiv \Gamma_{tot~(\pi^{K~0\prime}_{8~B})}/2
\end{eqnarray}

\noindent where the running of quark masses with energy is neglected, as well as 
phase-space factors differentiating scalar from pseudoscalar meson decays  which are
only important near the SM-fermion pair-production thresholds, assuming that mirror meson masses are not in that regime.
Moreover,  the total width of a meson has been defined above as
 approximately double its dominant decay width, 
in an effort to report later conservative order-of-magnitude cross-section estimates incorporating
 roughly not only theoretical but also experimental inefficiencies, uncertainties and cuts.

Another process of potential interest is the decay of $\sigma^K_B$ to a pair of group-``A" pseudoscalar mesons.
Assuming that the relevant meson decay amplitude is proportional to $g_{BA} (M_A/M_B) \sim r^{-2}$
where $g_{BA}$ is the effective $\sigma^K_B {\bar \pi^K_A} \pi^K_A$ meson coupling, 
an order-of-magnitude estimate for the decay width to a generic pair belonging to group-``A" mesons gives
\begin{equation}
\Gamma(\sigma^K_B \longrightarrow {\bar \pi^K_A} \pi^K_A) \approx \frac{g_{BA}^2M_A^2}{M_B^2}M_{\sigma^K_B}.
\end{equation}

\noindent 
There are two subgroups of group-``A" mesons, and each of these includes eight charged and eight
neutral color-octet pairs, twelve color-triplet pairs, as well as one charged and one neutral color-singlet pair. 
Assuming that each of these gives roughly the same decay amplitude,
an estimate of the total decay width of $\sigma_{B}^K$ to group-``A" mirror pseudoscalar mesons gives
\begin{equation}
\Gamma_{tot}(\sigma^K_B \longrightarrow {\bar \pi^K_A} \pi^K_A) \approx \frac{60g_{BA}^2M_A^2}{M_B^2}M_{\sigma^K_B}
\approx \frac{M_{\sigma^K_B}}{18}=32-64 {~\rm GeV},
\end{equation}

\noindent which should be added to the top-antitop quark decay width
for a correct order-of-magnitude estimate of the total
$\sigma^K_B$ decay width. Other large interesting classes of decays consist
 of group-``B" mirror-pion decays to three group-``A" mirror pions or CP-violating
decays to a pair of group-``A" mirror pions, but their
study exceeds the purposes of the present work. Extending techniques used to study QCD mesons in the present case
 in order to produce more reliable results could go along the lines of \cite{Hooft}.

Next come mirror-meson decay widths to two bosons mediated by loop diagrams \cite{Ellis}.
 Although factors depending on the size of the new gauge group lead to an enhancement of
the relative production mechanism in usual technicolor models, in katoptron models their effect is questionable due
to the breaking of the SU(3) mirror-family group \cite{TriantapIJMPA}. Moreover, due to a cancellation between
mirror leptons and mirror quarks, mirror pions do not decay to $W^{+}W^{-}$ and their decays to $Z^0 \gamma$ and
$Z^0 Z^0$ are suppressed. QCD corrections to diagrams involving gluons are neglected here
since we are mainly interested in 
the general qualitative features of the model. Some of the most interesting processes are listed
below:
\begin{eqnarray}
\Gamma (\sigma_{A,B}^{K} \longrightarrow gg) &\sim&
\frac{c_{\sigma g~A,B}\alpha_{s~A,B}^2(M_{\sigma_{A,B}^{K}}) M^3_{\sigma_{A,B}^{K}}}{216\pi^3 v^2} 
 \nonumber \\
\Gamma (\pi_{A,B}^{K~0\prime} \longrightarrow gg) &\sim&
\frac{c_{\pi g~A,B}\alpha_{s~A,B}^2(M_{\pi_{A,B}^{K~0\prime}}) M^3_{\pi_{A,B}^{K~0\prime}}}{96\pi^3 v^2}
\nonumber \\
\Gamma (\pi_{8~A,B}^{K~0\prime} \longrightarrow gg) &\sim& 
\frac{5 c_{\pi g~A,B}\alpha_{s~A,B}^2(M_{\pi_{8~A,B}^{K~0\prime}}) 
M^3_{\pi_{8~A,B}^{K~0\prime}}}{384\pi^3 v^2}
 \nonumber \\
\Gamma (\sigma_{~A,B}^{K} \longrightarrow \gamma\gamma) &\sim&
 \frac{c_{\sigma\gamma~A,B}\alpha^2(M_{\sigma_{A,B}^{K}}) M^3_{\sigma_{~A,B}^{K}}}{972\pi^3 v^2}
  \nonumber \\
\Gamma (\pi_{~A,B}^{K~0 \prime} \longrightarrow \gamma\gamma) &\sim&
 \frac{c_{\pi\gamma~A,B}\alpha^2(M_{\pi_{A,B}^{K~0\prime}}) M^3_{\pi_{~A,B}^{K~1~0 \prime}}}{432\pi^3 v^2}
\end{eqnarray}

\noindent where the first and second terms of each subscript correspond to group-A and group-B mesons respectively, 
katoptrons are assumed to be much heavier than mirror mesons, $\alpha$ 
and $\alpha_s$ are the usual electromagnetic and QCD structure constants and the prefactors
$c_{\sigma g~A,B}, c_{\pi g~A,B}, c_{\sigma\gamma~A,B}, c_{\pi\gamma~A,B}$ codify the interference from different sources discussed below. Setting these prefactors equal to unity corresponds to triangle diagrams involving a single heavy katoptron generation.

Note that, in the limit of very large top-quark mass with respect to group-``A" mirror-meson masses, the expression for the two-boson meson decay width $\Gamma^{top}$
 involving a top-quark triangle diagram yields
\begin{eqnarray}
\frac{c_{\sigma g~A}\Gamma^{top} (\sigma_{A}^{K} \longrightarrow gg)}{\Gamma(\sigma_A^{K} \longrightarrow gg)}&=&
\frac{c_{\pi g~A}\Gamma^{top}(\pi_{A}^{K} \longrightarrow gg)}{\Gamma(\pi_A^{K} \longrightarrow gg)}=3 \nonumber \\
\frac{c_{\sigma\gamma~A}\Gamma^{top}(\sigma_{A}^{K} \longrightarrow \gamma\gamma)}{\Gamma (\sigma_A^{K} \longrightarrow \gamma\gamma)}&=&
\frac{c_{\pi\gamma~A}\Gamma^{top}(\pi_{A}^{K} \longrightarrow 
\gamma\gamma)}{\Gamma(\pi_A^{K}\longrightarrow\gamma\gamma)}= 12,
\end{eqnarray}

\noindent relations which are partially due to the normalization of the generator of the $SU(8)$ chiral symmetry corresponding to the mirror mesons and to the large splitting of top and bottom quark masses.
Moreover, assuming that scalar and pseudoscalar mesons have a common mass $M_{A,B}$, 
one has the following relations:
\begin{equation}
\frac{c_{\sigma\gamma~A,B}\Gamma(\sigma_{A,B}^{K} \longrightarrow gg)}{c_{\sigma g~A,B}\Gamma(\sigma_{A,B}^{K} \longrightarrow \gamma\gamma)}= 
\frac{c_{\pi\gamma~A,B}\Gamma(\pi_{A,B}^{K} \longrightarrow gg)}{c_{\pi g~A,B}\Gamma(\pi_{A,B}^{K} 
\longrightarrow \gamma\gamma)}=
\frac{2\alpha_{s~A,B}^2(M_{A,B})}{ 9\alpha_{A,B}^2(M_{A,B}) }
\end{equation}
\noindent while, again in the limit of large fermion masses, one has
\begin{equation}
\frac{c_{\sigma g~A,B}\Gamma(\pi_{A,B}^{K} \longrightarrow gg)}{c_{\pi g~A,B}\Gamma(\sigma_{A,B}^{K}
 \longrightarrow gg)}= 
\frac{c_{\sigma\gamma~A,B}\Gamma(\pi_{A,B}^{K} \longrightarrow \gamma\gamma)}{c_{\pi\gamma~A,B}\Gamma(\sigma_{A,B}^{K} 
\longrightarrow \gamma\gamma)}=9/4.
\end{equation}

Regarding the  interference of various sources participating in the loop diagrams involved in the mirror-meson couplings with
two gauge bosons, the following remarks are in order: 

$c_{\sigma g~A,~\pi g~A}$: 
First, each of the two kinds of group-``A" mirror mesons are produced by 
gluon-fusion triangle diagrams of a top quark interfering with $\delta=1$ or 2 mirror fermion generations,
neglecting contributions from lighter SM fermions. 
In case $\delta=1$, these two kinds of mirror mesons are distinct and should have comparable but 
different masses. 
However, in case $\delta=2$, one expects to detect only one kind of group-``A" mesons, followed by a larger 
enhancement of the relevant production and decay cross-sections. This might explain the slight excess in total 
Higgs production cross-section \cite{CERNexcess}. However, it is important to stress at this point that
the eventual self-breaking of $SU(2)_K$ might be the cause of a 
- partially at least- destructive interference between
the contributions coming from
the two lighter katoptron generations, damping thus the final enhancement effect.
Moreover, note the assumption that group-``A" mirror mesons are taken to
have zero tree-level couplings to the heaviest katoptron generation.

$c_{\sigma g~B}$: On the other hand, group-``B"
mirror scalar mesons are produced by 
gluon-fusion loop 
diagrams involving group-``A" mirror mesons interfering  with only the heaviest mirror fermion generation,
neglecting all lighter katoptron and SM-fermion contributions. The reason why
the relative lightness of group-``A" mirror mesons does not lead to their decoupling,
contrary to what happens with light fermions, is 
analyzed in detail in \cite{Marciano}.

$c_{\pi g~B}$: The same
is true for pseudoscalar group-``B" mirror mesons if there is CP-violation,
while, if CP symmetry is conserved, only the heaviest katoptron generation contributes to group-``B" mirror-meson production. Contributions of pseudoscalars which are heavier than the decaying mirror mesons of either group 
are assumed to decouple, their influence being restrained to the decay
constants of group-``A" mirror mesons for instance, and are neglected \cite{Boston}.

$c_{\sigma \gamma~A}$: 
Furthermore, each of the two kinds of scalar group-``A" mirror-meson decay
 to two photons proceed via
interfering loop diagrams of $\delta=1,2$
katoptron generations with a top quark and a $W$ gauge boson, neglecting all lighter SM fermions and
noting that the contribution of the katoptron generations is 
relatively small and is not expected to lead to large deviations from the corresponding standard Higgs decay.

$c_{\sigma \gamma~B}$:  On the other hand, group-``B"
mirror scalar mesons decay to two photons via loops involving the heaviest katoptron generation interfering with a $W$
boson and group-``A" mirror mesons, neglecting all lighter SM and mirror fermion generations. 

$c_{\pi \gamma~A}$: Last, each of the two kinds of 
group-``A" pseudoscalar mirror mesons decay to two photons via
 interfering loops of the top quark with $\delta$ katoptron generations, again neglecting lighter SM fermions, 
while 

$c_{\pi \gamma~B}$: group-``B" pseudoscalar
 diphoton mirror-meson 
decays proceed via loop diagrams involving the heaviest katoptron generation, 
possibly interfering with loops of group-``A" mirror mesons if
CP symmetry is violated, neglecting all lighter SM and mirror fermion generations.

The above remarks lead to the following definitions:
\begin{eqnarray}
c_{\sigma g~A} &=& c_{\pi g~A} = |\delta+e^{i\theta_{g~A}}\sqrt 3|^2  \nonumber \\
c_{\sigma \gamma~A} &=& |\delta+e^{i\theta_{\gamma~A}}2\sqrt{3}(1+I^W_A)|^2  \nonumber \\
c_{\pi \gamma~A} &=& |\delta+e^{i\theta_{\gamma~A}}2\sqrt{3}|^2  \nonumber \\
c_{\sigma g~B} &=& |1+e^{i\theta_{g~B}}\sqrt{3}I^{g}|^2  \nonumber \\
c_{\pi g~B} &=& |1+\epsilon_{g}e^{i\theta_{g~B}}\sqrt{3}I^{g}|^2  \nonumber \\
c_{\sigma \gamma~B} &=& 
|1+2\sqrt{3}(e^{i\theta_{\sigma\gamma}}I^{\gamma}+e^{i\theta_{\gamma~B}} I^W_B)|^2  \nonumber \\
c_{\pi\gamma~B}&=& |1+\epsilon_{\gamma}e^{i\theta_{\pi\gamma}}2\sqrt{3}I^{\gamma}|^2
\end{eqnarray}

\noindent with $\theta_{g~A,B}, \theta_{\gamma~A,B}, \theta_{\sigma\gamma,\pi\gamma}$
  interference phases between various sources, $\epsilon_{g,\gamma}$ CP-violation parameters and 
$I^{g,\gamma}$ loop 
contributions of group-``A" mirror pseudoscalar mesons to gluon-gluon fusion and 
diphoton decay amplitudes respectively, and $I^{W}$ the W-boson contribution to mirror scalar meson decay amplitudes.
Following the expressions given in 
\cite{Marciano} for the W-boson contribution to the two-photon decay, one finds
 $I^W_A \approx -4.7$, which is quite larger than fermionic contributions, while $I^W_B \approx -1.12$.
(The quantity $I^W_B$ might be further suppressed by a
 factor of $g_{BA}\sim1/r$ since the W gets its mass by 
``eating" group-``A" mirror mesons, but as  will become clear shortly this does not 
affect the final result significantly since $I^{\gamma}$ is quite larger anyway.)

Note that the effect of these parameters might obviously alter the production rates and
branching ratios of various processes in a way that they may be distinguished from
 the ones expected by their SM-type values. Furthermore, even though one could have used the gauged
Wess-Zumino-Witten
formalism \cite{WZW} to parametrize the mirror-meson diboson decay 
widths (see \cite{WZWexamples} for instance), 
a detailed estimation approach is chosen  instead in order to have a closer control on the final results. Last,
 the possibility is noted that CP violation leads either to the mixing  of scalar
and pseudoscalar resonances if their are mass-degenerate, or to non-zero couplings of pseudoscalar mirror mesons to the $W$ bosons \cite{Holdom} altering thus $c_{\pi\gamma~A,B}$ further.

In more detail, contributions of color-singlet $I^{\gamma}_0$, 
leptoquark $I^{g,\gamma}_3$ and color-octet mirror mesons $I^{g,\gamma}_8$
 to loop diagrams involving either two gluons or two photons, assuming that group-``B" mirror mesons are
much heavier than group-``A" mirror mesons, and recalling that the quadratic Casimir invariants of the fundamental
and adjoint complex representation of SU(3) are equal to 1 and 6 respectively,
 lead to the
following estimates:
\begin{eqnarray}
I^g &=& I^g_3+I^g_8 \sim g_{BA}(2\cdot1\cdot8+2\cdot6\cdot4) = 64/r\nonumber \\
I^{\gamma} &=&I^{\gamma}_0+ I^{\gamma}_3+I^{\gamma}_8 \sim  g_{BA}\Big\{2\cdot1+2\cdot3\left((1/3)^2+2(2/3)^2+(5/3)^2\right)+2\cdot8\cdot1\Big\}=\frac{122}{3r}
\end{eqnarray}

\noindent where the expressions above imply  common $\sigma^K_B-\pi^K_A\pi_A^K$ and
$\pi_B^K-\pi^K_A\pi_A^K$ effective couplings $g_{BA} \sim r^{-1}$,
implying in parallel that the two group-``A" mirror-meson contributions dominate group ``B" mirror-meson couplings to two gauge bosons.

Furthermore, the running of the gauge couplings entering the decay formulas is given by approximating 
$\alpha(p) \approx  \alpha(M_Z)$ and
\begin{equation}
\alpha_s(p)=\left\{\begin{array}{cc}
\alpha_{s~A}\equiv\left[\alpha_s^{-1}(M_Z)+\frac{21}{6\pi}\ln{(p/M_Z)}\right]^{-1}
{\rm ~for ~group~ ``A" ~mesons}& \\
\alpha_{s~B}\equiv\left[\alpha_s^{-1}(M_Z)+\frac{21}{6\pi}\ln{(p/(rM_Z))}+\frac{13}{6\pi}\ln{(r)}\right]^{-1}
{\rm ~for ~group~ ``B" ~mesons}&
\end{array} \right.
\end{equation}
\noindent where $\alpha(M_Z) \approx 1/129$ and $\alpha_s(M_Z) \approx 1/8.5$, neglecting the renormalization
of $\alpha$ at energies higher than the mass  $M_Z$ of the $Z$ boson
due to its relatively small effect and the effect of colored mirror mesons on the running
of the strong coupling, while taking into account the decoupling of group-``B" katoptrons at
energies on the order of group-``A" mirror-meson masses.

\subsection{Computing the Cross-Sections}
Analytic expressions for several
 interesting cross-sections are listed below, since all the needed
 ingredients are now at hand.
Cuts to reduce the relevant QCD background, when applicable, would roughly decrease
them by half. The role of the Higgs is played by $\sigma^K_A$, where one only expects a slight production enhancement from
katoptrons participating in the gluon-fusion triangle diagram. One should
expect the detection of neutral pseudoscalars as mixtures of $\pi^{K~0 \prime}_{A,B}$ with $\pi^{K~a~0}_{A,B}$
at masses close to the corresponding charged pseudoscalars. However, lack of observed decays of the top quark to these 
charged pseudoscalars indicates that, in case they are distinct particles, they should be all heavier than about 170 GeV. 

The  relevant expressions are given by:
\begin{eqnarray}
\sigma({\bar p} p \longrightarrow \sigma^{K}_{A} \longrightarrow {\bar \tau} \tau) &=& 
{\cal L}(M_{\sigma^K_A})
\frac{c_{\sigma g~A}}{\pi}\left(\frac{\alpha_{s~A}(M_{\sigma^K_A})M_{\sigma^K_A}m_{\tau}}{36\pi vm_b}\right)^2
\nonumber \\
\sigma({\bar p} p \longrightarrow \sigma^{K}_{A} \longrightarrow {\bar b} b) &=& {\cal L}(M_{\sigma^K_A})
\frac{3c_{\sigma g~A}
}{\pi}\left(\frac{\alpha_{s~A}(M_{\sigma^K_A})M_{\sigma^K_A}}{36\pi v}\right)^2
\nonumber \\
\sigma({\bar p} p \longrightarrow \sigma^{K}_{A} \longrightarrow \gamma \gamma) &=&
 {\cal L}(M_{\sigma^K_A})
\frac{3c_{\sigma g~A}c_{\sigma\gamma~A}}{2\pi}
\left(\frac{\alpha_{s~A}(M_{\sigma^{K}_{A}})
\alpha(M_{\sigma^{K}_{A}}) M^2_{\sigma^{K}_A}}{486\pi^2 vm_b}\right)^2
\nonumber 
\end{eqnarray}
\begin{eqnarray}
\sigma({\bar p} p \longrightarrow \pi^{K~0 \prime}_A \longrightarrow {\bar \tau} \tau) &=& 
{\cal L}(M_{\pi^{K~0 \prime}_A})
\frac{c_{\pi g~A}
}{\pi}\left(\frac{\alpha_{s~A}(M_{\pi^{K~0 \prime}_A})M_{\pi^{K~0 \prime}_A}m_{\tau}}{24\pi vm_b}\right)^2
\nonumber \\
\sigma({\bar p} p \longrightarrow \pi^{K~0 \prime}_A \longrightarrow {\bar b} b) &=& {\cal L}(M_{\pi^{K~0 \prime}_A})
\frac{3c_{\pi g~A}
}{\pi}\left(\frac{\alpha_{s~A}(M_{\pi^{K~0 \prime}_A})M_{\pi^{K~0 \prime}_A}}{24\pi v}\right)^2
\nonumber \\
\sigma({\bar p} p \longrightarrow \pi^{K~0 \prime}_A \longrightarrow \gamma \gamma) &=& {\cal L}(M_{ \pi^{K~0 \prime}_A})
\frac{3c_{\pi g~A}c_{\pi\gamma~A}}{2\pi}\left(\frac{\alpha_{s~A}(M_{\pi^{K~0 \prime}_A})\alpha(M_{\pi^{K~0 \prime}_A}) M^2_{\pi^{K~0 \prime}_A}}{216\pi^2 vm_b}\right)^2
\nonumber 
\end{eqnarray}
\begin{eqnarray}
\sigma({\bar p} p \longrightarrow \sigma^{K}_{B} \longrightarrow {\bar \tau} \tau) &=& {\cal L}(M_{\sigma^K_B})
\frac{c_{\sigma g~B}}{\pi}\left(\frac{\alpha_{s~B}(M_{\sigma^K_B})M_{\sigma^K_B}m_{\tau}}{36\pi vm_t}\right)^2
\nonumber \\
\sigma({\bar p} p \longrightarrow \sigma^{K}_{B} \longrightarrow {\bar t} t) &=& {\cal L}(M_{\sigma^K_B})
\frac{3c_{\sigma g~B}}{\pi}\left(\frac{\alpha_{s~B}(M_{\sigma^K_B})M_{\sigma^K_B}}{36\pi v}\right)^2
\nonumber \\
\sigma({\bar p} p \longrightarrow \sigma^{K}_{B} \longrightarrow \gamma \gamma) &=& {\cal L}(M_{\sigma^K_B})
\frac{3c_{\sigma g~B}c_{\sigma\gamma~B}}{2\pi}
\left(\frac{\alpha_{s~B}(M_{\sigma^{K}_{B}})\alpha(M_{\sigma^{K}_{B}}) M^2_{\sigma^{K}_B}}{486\pi^2 vm_t}\right)^2
\nonumber 
\end{eqnarray}
\begin{eqnarray}
\sigma({\bar p} p \longrightarrow \pi^{K~0 \prime}_B \longrightarrow {\bar \tau} \tau) &=& {\cal L}(M_{ \pi^{K~0 \prime}_B})
\frac{
c_{\pi g~B}}{\pi}\left(\frac{\alpha_{s~B}(M_{\pi^{K~0 \prime}_B})M_{\pi^{K~0 \prime}_B}m_{\tau}}{24\pi vm_t}\right)^2
\nonumber \\
\sigma({\bar p} p \longrightarrow \pi^{K~0 \prime}_B \longrightarrow {\bar t} t) &=& 
{\cal L}(M_{\pi^{K~0 \prime}_B})
\frac{3c_{\pi g~B}}{\pi}\left(\frac{\alpha_{s~B}(M_{\pi^{K~0\prime}_B})M_{\pi^{~0\prime}_B}}{24\pi v}\right)^2
\nonumber \\
\sigma({\bar p} p \longrightarrow  \pi^{K~0 \prime}_B \longrightarrow \gamma \gamma) &=& {\cal L}(M_{\pi^{K~0 \prime}_B})
\frac{3c_{\pi g~B}c_{\pi\gamma~B}}{2\pi}\left(\frac{\alpha_{s~B}(M_{\pi^{K~0 \prime}_B})\alpha(M_{\pi^{K~0 \prime}_B}) M^2_{\pi^{K~0 \prime}_B}}{216\pi^2 vm_t}\right)^2
\end{eqnarray}

\begin{figure}
\centering
\includegraphics
[angle=0,width=17cm]
{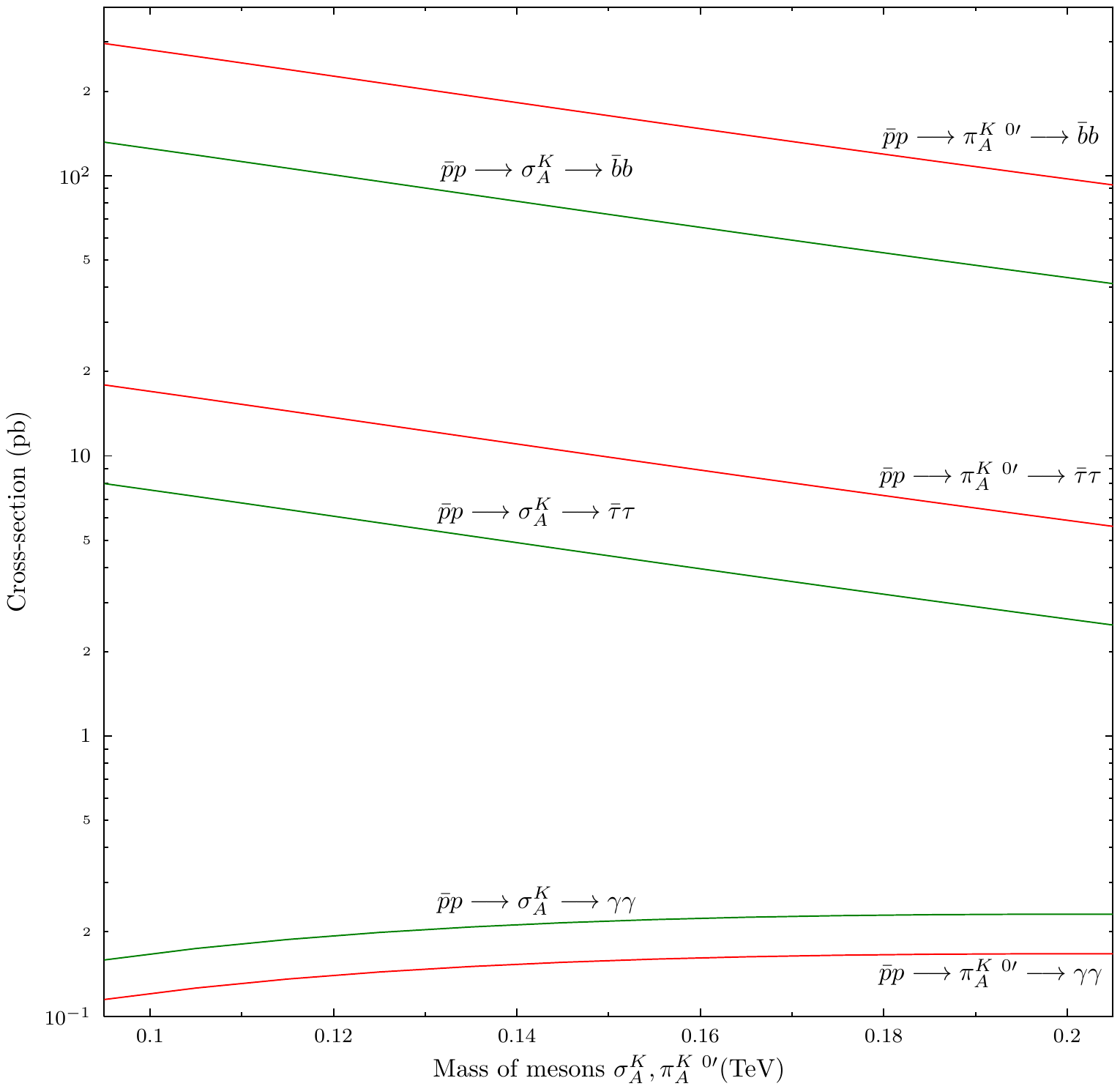}
\vspace{-5cm}
\caption{Cross-section estimates of group-``A" spin-0 neutral color-singlet mirror-meson processes. 
Green lines correspond to scalar and red lines
to pseudoscalar processes.}
\end{figure}

\begin{figure}
\centering
\includegraphics
[angle=0,width=17cm]
{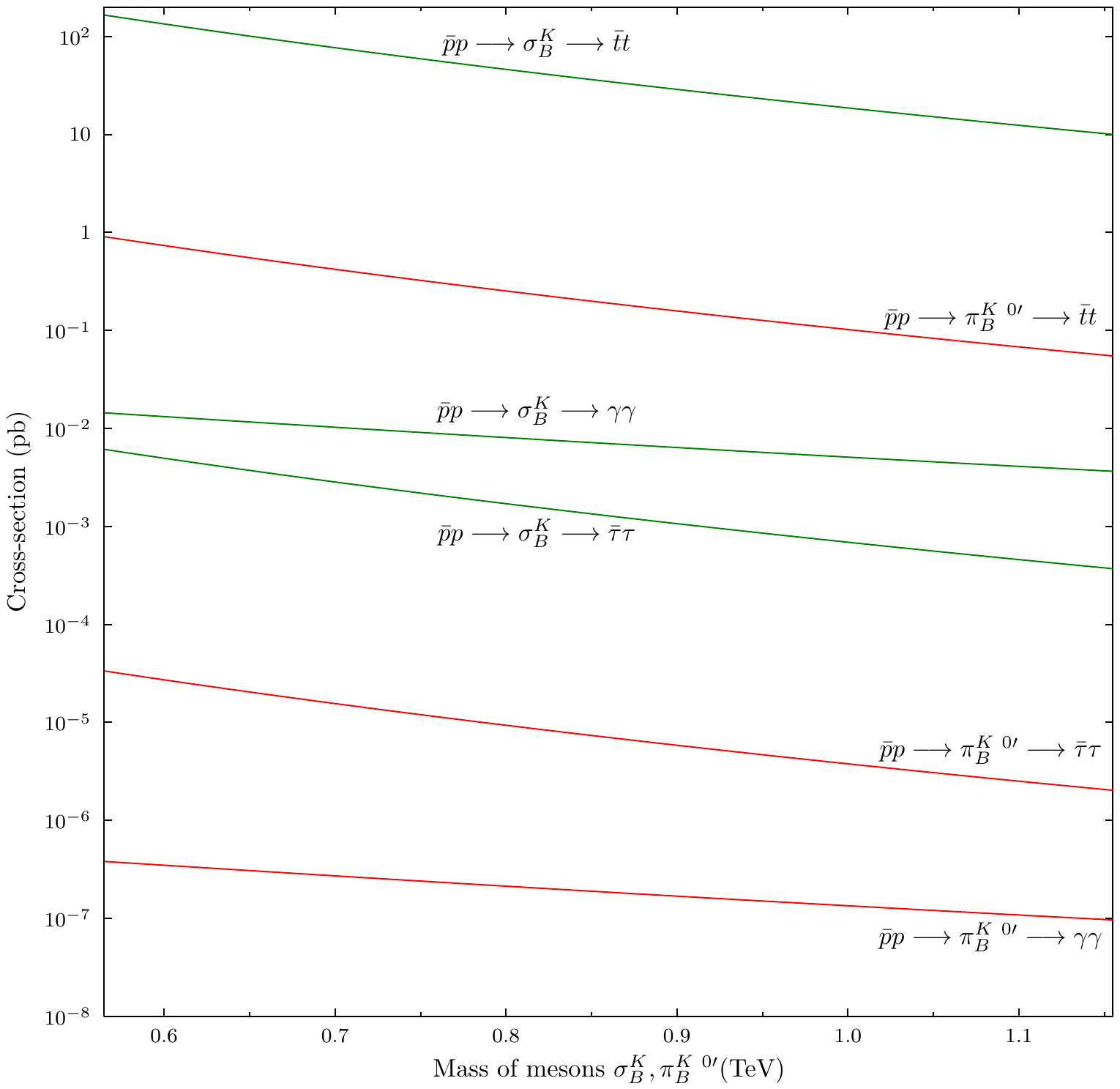}
\vspace{-5cm}
\caption{Cross-section estimates of group-``B"  spin-0 neutral color-singlet mirror-meson processes.
Green lines correspond to scalar and red lines
to pseudoscalar processes.}
\end{figure}

\noindent 
The corresponding results are visualized in Figs. 2 and 3, assuming all interference phases and CP-violation parameters 
are zero and $\delta=1$. Scalar and pseudoscalar mirror-meson decays to gluon pairs might also be of interest if they can be
distinguished from QCD background, but they are left for future work since their study is 
better suited for a lepton collider. 

Scalar-meson $\sigma_{A}^{K}$ processes in Fig. 2 should agree with the Higgs-like particle
already observed at the LHC. Spin-0 mirror-meson decays to fermions are Higgs-like so they are not expected to create problems
when compared to SM expectations. 
Furthermore, loop contributions stemming from top-quarks and W-bosons to spin-0 meson decays into
dibosons of the electroweak sector and diphotons in particular
are quite larger than katoptron ones, rendering thus
deviations from SM expectations manageable, especially if non-zero interference phases are introduced.
Moreover, it is obvious that one should seek pseudoscalar group-``A" mirror 
meson decays to bottom quark and $\tau$ pairs which dwarf the corresponding  relatively very small diphoton cross-sections due to the absence of the $W$-boson contribution. 
Stringent rapidity or high transverse momentum cuts are needed to minimize the 
relevant QCD background. 

The case where group-``A" mirror-meson decays to top-antitop quark pairs are kinematically allowed is
not studied in this work although it is conceivable,
but is is obvious that it would lead to much larger cross-sections than the ones presented here.
On the contrary, pseudoscalar group-``B" mirror 
meson decays to top quark pairs are not as important as the corresponding scalar meson decays due to the absence of 
enhancement of their production via gluon fusion, unless the CP violating parameters $\epsilon_{g,\gamma}$ turn out to be non-negligible. 

Cross-sections of group-``B" mirror scalar mesons are more important than
the corresponding pseudoscalar ones due to group-``A" mirror-meson loop contributions, 
unless there exist large CP-violating couplings.
In particular, it should be noted that the enhancement of group-``B" scalar-meson production and diphoton decay via
loops involving group-``A" mirror mesons renders the process ${\bar p} p \longrightarrow \sigma^{K}_{B} \longrightarrow \gamma \gamma$ particularly interesting because it is easily distinguished from background,
and might be linked to the diphoton excesses at invariant masses close to 750 GeV 
in the ATLAS and CMS CERN experiments \cite{750} being on the order of 10 fb and corresponding to a mirror meson
exhibiting a similar
decay width. 

Given the mass (125 GeV) of the Higgs-like particle already discovered 
corresponding to $\sigma_A^K$  in our case, confirmation of these excesses 
and identification of their source with $\sigma^K_B$ would imply a
intra-generational katoptron hierarchy close to $r \sim 6$. This is remarkably close to the
rough estimate $r \approx 5.75$ of the previous section, taking into account the non-perturbative dynamics involved and the roughness of the calculation.
 The strength of the signal can be traced to the large number of group-``A" mirror pions
which are coupled to $\sigma^K_{B}$. However, non-perturbative dynamics
do not allow a precise determination of the magnitude of the coupling between the
various groups of mirror mesons. This
implies that it is possible that the actual signals relevant to these mirror-meson decays turn out
to be finally considerably weaker without creating serious problems to the viability of the model.

Near-degeneracy of scalar and pseudoscalar mirror-meson masses might
also lead to a signal which is a combination of several resonances. Moreover, larger effective and/or CP-violating 
couplings than the ones considered here might be able in principle to explain these excesses,
alternatively though less likely, in terms of $\pi^{K~0\prime}_B$ or $\eta^{K~\prime}_B$.
Furthermore, consistency
of the theory implies obviously that one should also expect a quite strong signal corresponding to 
${\bar p} p \longrightarrow \sigma^{K}_{B} \longrightarrow {\bar t} t$, which should become clearer from QCD background by the cuts mentioned above. 

Other processes of interest which are not followed by corresponding diphoton decays are
pseudoscalar color-octet mirror-meson decays:
\begin{eqnarray}
\sigma({\bar p} p \longrightarrow \pi^{K~0 \prime}_{8~A} \longrightarrow {\bar b} b) &=& {\cal L}(M_{\pi^{K~0 \prime}_{8~A}})
\frac{15 c_{\pi g~A}}{\pi}
\left(\frac{\alpha_{s~A}(M_{\pi^{K~0 \prime}_{8~A}})M_{\pi^{K~0 \prime}_{8~A}}}{24\pi v}\right)^2
\nonumber \\
\sigma({\bar p} p \longrightarrow \pi^{K~0 \prime}_{8~B} \longrightarrow {\bar t} t) &=&{\cal L}(M_{\pi^{K~0\prime}_{8~B}})
\frac{15c_{\pi g~B}}{\pi}\left(\frac{\alpha_{s~B}(M_{\pi^{K~0\prime}_{8~B}})M_{\pi^{K~0\prime}_{8~B}}}{24\pi v}\right)^2
\end{eqnarray}

\noindent Color factors increase the relevant cross-sections compared to the ones involving color-singlet mirror mesons. As before, we report results in Figs. 4 and 5
assuming zero interference phases and CP violation, while taking $\delta=1$.
Moreover, decays of $\pi^{K~0 \prime}_{8~A,B}$ to two gluons might in principle be of interest if they can be distinguished 
from QCD background by appropriate cuts but they are left for future work 
relevant to lepton colliders. Note moreover that the
case where the decay $\pi^{K~0 \prime}_{8~A} \longrightarrow {\bar t} t$ is kinematically
allowed cannot be totally excluded, 
something that would lead to a much larger relevant cross-section.

Production of pairs of mirror pseudoscalar mesons is studied next. The ones studied here are
 either color-octets or leptoquarks.
These are mainly produced by gluon fusion, but the relevant cross-sections can be enhanced by intermediate
color-octet vector mirror mesons $\rho^{K~0 \prime}_{8~A,B}$. 
Nevertheless, the mirror meson mass estimates in the previous section imply that the decays
 $\rho^{K~0 \prime}_{8~A,B}\longrightarrow \pi^{K~+}_{8~A,B} ~ \pi^{K~-}_{8~A,B}$ are likely
not allowed kinematically. Even when they are, due to vector-meson-dominance arguments
 their contribution to the total cross-section is expected to be of the same order of magnitude
as the QCD one \cite{Terning}. In the following, it is assumed that the production cross-section
 of two pseudoscalar color-octets is dominated by QCD via gluon fusion ignoring the mirror vector-meson contribution, 
and it is given by  \cite{EHLQ},\cite{Manohar}
\begin{equation}
\sigma({\bar p} p \longrightarrow
\pi^{K~+}_{8~A,B} ~ \pi^{K~-}_{8~A,B} \longrightarrow {\bar b}t+ {\bar t} b) \approx
{\cal L}(2M_{\pi^{K~\pm}_{8~A,B}})27c_w\pi
\left(\frac{\alpha_s(2M_{\pi^{K~\pm}_{8~A,B}})}{32}\right)^2.
\end{equation}

\noindent 
The production
cross-section above is integrated over a narrow invariant-mass bin 
$2M_{\pi^{K~\pm}_{8~A,B}}(1+ w)$ over the mirror-meson pair-production  mass threshold.
To account for this, a parameter $c_w$ is introduced, given by
 \begin{equation}
c_w \approx \frac{2(2w)^{3/2}}{3}, 
\end{equation}
neglecting the running of the gluon distribution functions
over that bin. For the results presented, 
the choice $w=5.6\%$ has been made so that $c_w=2.5\%$.
Moreover, branching ratios of the dominant decay studied here 
are taken to be equal to 1/2 for both color-octet mesons, reducing the final cross-section 
further by a factor of 4. 

The final quarks produced are generally
not collinear with their anti-particles, leading to signals above a certain mass threshold  involving top and bottom quarks
which may in principle be distinguished from QCD background via appropriate acollinearity and rapidity cuts. This property is not shared by the production and decays of pairs of
neutral color-octet mirror pseudoscalars, which is expected to lead to
an enhancement of top-antitop quark and gluon-pair production at energies
above their production threshold. However, the significant QCD background to these
processes necessitates a more detailed study which is left for work relevant to future
lepton colliders.

\begin{figure}
\centering
\includegraphics
[angle=0,width=20cm]
{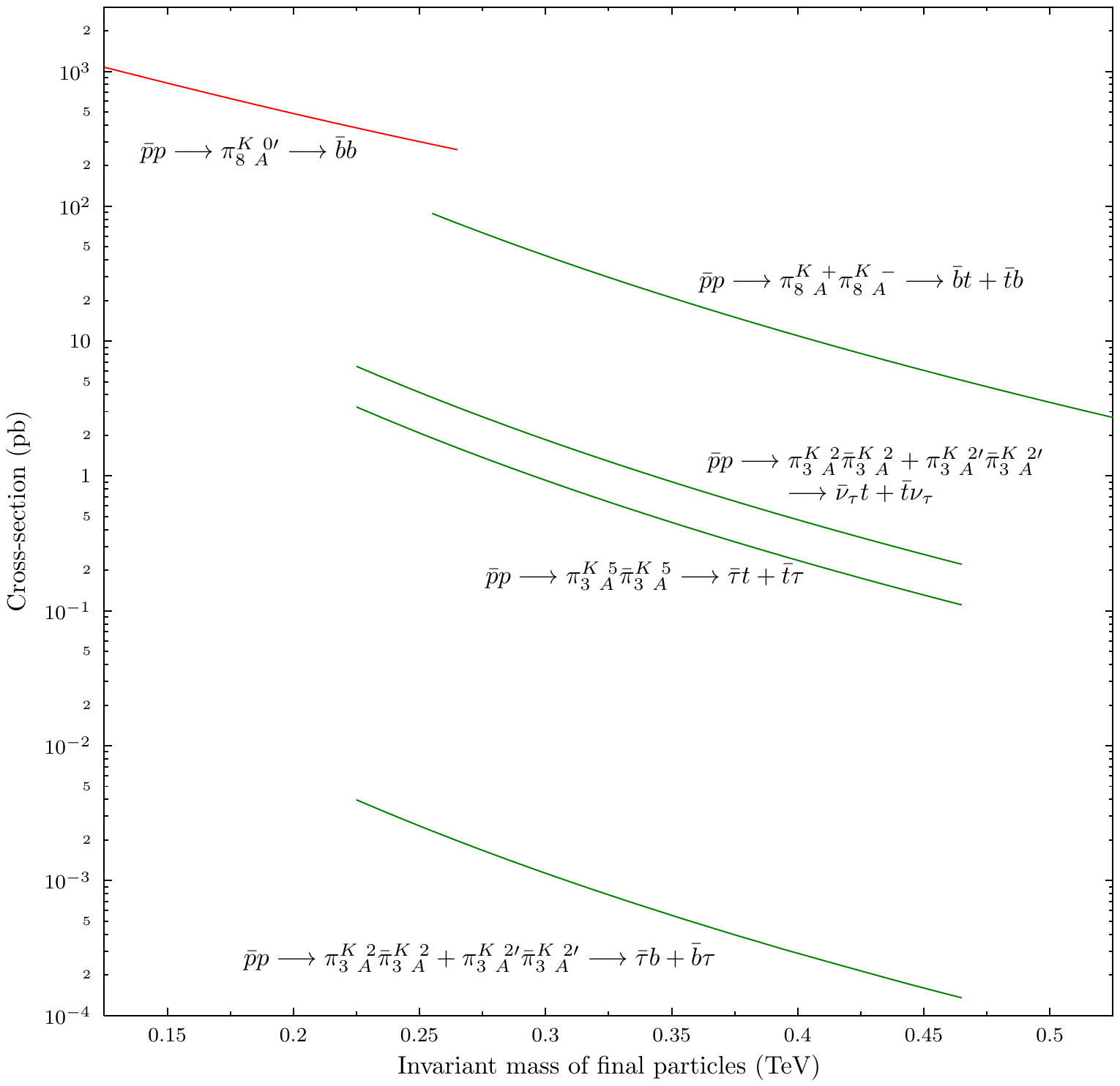}
\caption{Cross-section estimates of group-``A" mirror-meson processes.
Green lines correspond to production of two colored mirror mesons while the red line
to one pseudoscalar octet.}
\end{figure}

\begin{figure}
\centering
\includegraphics
[angle=0,width=20cm]
{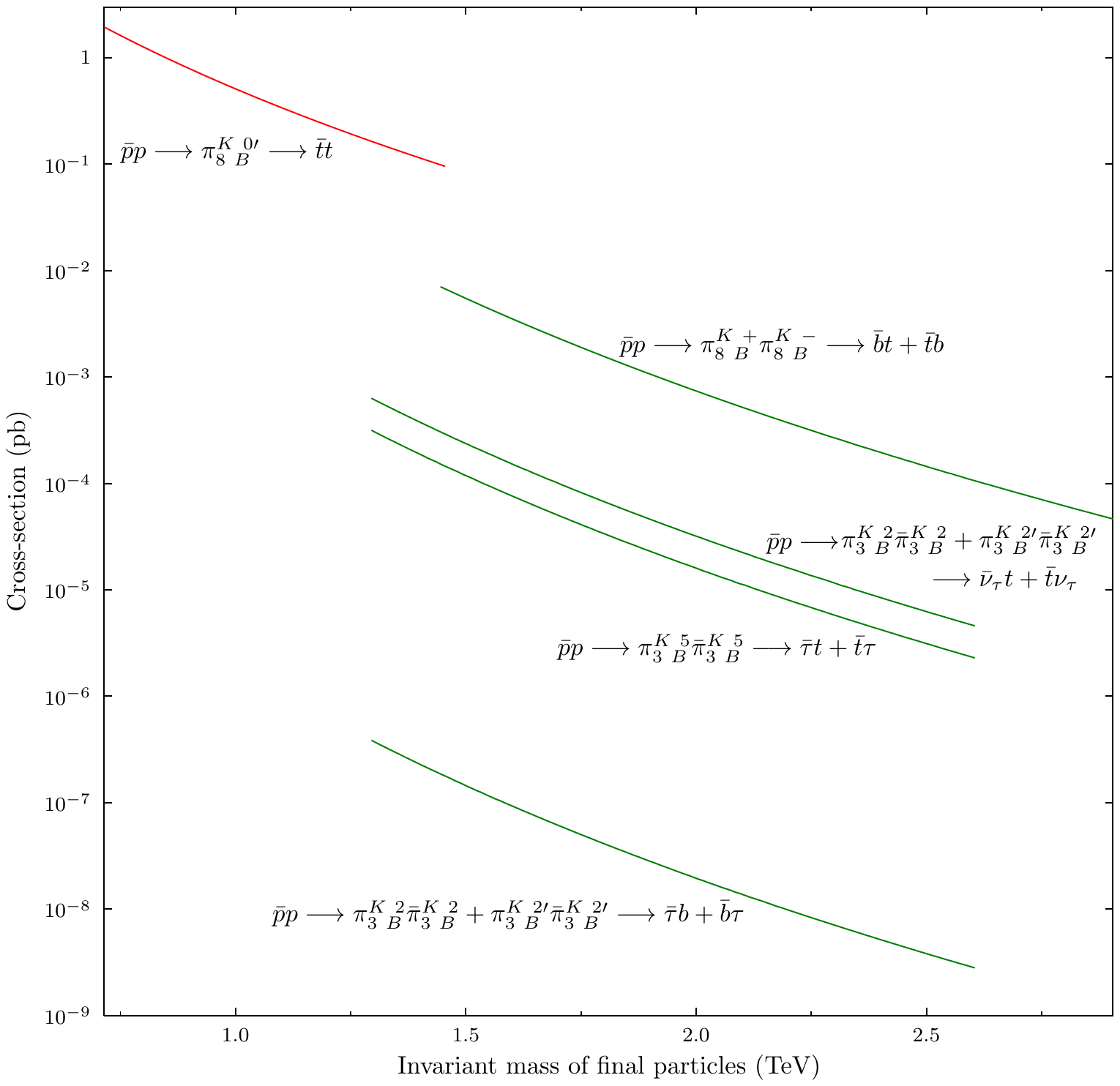}
\caption{Cross-section estimates of group-``B" mirror-meson processes.
Green lines correspond to production of two colored mirror mesons while the red line
to one pseudoscalar octet.}
\end{figure}

A similar analysis is followed
for the production cross-section of pairs of pseudoscalar leptoquarks for the sake of
simplicity, even  though the decays
$\rho^{K~0 \prime}_{8~A,B}\longrightarrow {\bar \pi^{K~1,2,2\prime,5}_{3~A,B}} ~ \pi^{K~1,2,2\prime,5}_{3~A,B}$
might be marginally allowed kinematically.  The ones of most interest due to the heaviness of the final fermions
 are given below, assuming that $\pi^{K~2}_{3~A,B}$ and $\pi^{K~2\prime}_{3~A,B}$ are mass-degenerate:
\vspace{-0.0mm}
\begin{eqnarray}
\sigma({\bar p} p \longrightarrow
\pi^{K~5}_{3~A,B} ~ {\bar \pi^{K~5}_{3~A,B}} &\longrightarrow & {\bar \tau}t + {\bar t} \tau) = 
\nonumber \\
 &=&{\cal L}(2M_{\pi^{K~5}_{3~A,B}})\frac{7c_w\pi}{12}
\left(\frac{\alpha_s(2M_{\pi^{K~5}_{3~A,B}})}{32}\right)^2 \nonumber \\
\sigma({\bar p} p\longrightarrow
\pi^{K~2}_{3~A,B} ~ {\bar \pi^{K~2}_{3~A,B}} +
\pi^{K~2\prime}_{3~A,B} ~ {\bar \pi^{K~2\prime}_{3~A,B}}&\longrightarrow & {\bar \nu_{\tau}}t + {\bar t} \nu_{\tau}
) = \nonumber \\
 &=&{\cal L}(2M_{\pi^{K~2}_{3~A,B}})\frac{7c_w\pi}{6}
\left(\frac{\alpha_s(2M_{\pi^{K~2}_{3~A,B}}) }{32}\right)^2\nonumber \\
\sigma({\bar p} p\longrightarrow
\pi^{K~2}_{3~A,B} ~ {\bar \pi^{K~2}_{3~A,B}} +
\pi^{K~2\prime}_{3~A,B} ~ {\bar \pi^{K~2\prime}_{3~A,B}}&\longrightarrow & {\bar \tau}b + {\bar b} \tau) =
\nonumber \\
 &=&{\cal L}(2M_{\pi^{K~2}_{3~A,B}})\frac{7c_w\pi}{6}
\left(\frac{\alpha_s(2M_{\pi^{K~2}_{3~A,B}}) m_b}{32m_t}\right)^2 
\nonumber \\
 &&
\end{eqnarray}

 \noindent where the dominant decay widths of each leptoquark are taken again to have branching ratios of 1/2. 
Acollinearity and rapidity cuts applied to the particle-antiparticle decay products are expected in this case also to
reduce the QCD background to these processes
significantly, offering interesting signals pointing to the existence of mirror leptoquarks above their production threshold.
Current experimental mass limits on leptoquarks decaying to a bottom quark and a $\tau$ can easily be circumvented since 
they do not take into account that the relevant decay cross-section in the
mirror-meson case is suppressed by a factor of $(m_b/m_t)^2 < 10^{-3}$.
The corresponding results can be visualized in Figs. 4 and 5.

Detection of vector color-octet mesons $\rho^{K~0 \prime}_{8~A,B}$, apart from a possible 
enhancement of color-octet pseudoscalar mirror-meson and mirror-leptoquark pair production, might alternatively come from their decays to two gluons, 
even though such signals might be hard to distinguish from QCD background  unless
one employs a leptonic collider. Furthermore,
of interest might be much rarer processes involving $\rho_3^K$ vector 
mesons and  heavier mirror hadrons possibly existing which are singlets under the mirror-family group
and involve three katoptrons (like mirror protons $p^K$ and mirror neutrons $n^K$), or 
even four and five katoptrons (analogous to QCD tetra- and penta-quarks). In any case, 
having introduced the
actors and shown a small act
 of the play inspired by the katoptron model, it is time to pose the important question regarding
 the way the stage is set for them.

\section{Implications}

It is clear that observation of mirror meson decays similar to the ones described above
at the RUNII in LHC would open new horizons in elementary particle Physics.
Apart from explaining the BEH mechanism, it might also offer useful insights related to the hierarchy problem, the
paradigm of unification of gauge interactions, and possibly quantum gravity.
In particular, the existence of mirror fermions appears naturally in a discretized version of spacetime
in a spirit similar to \cite{Kleinert}. Such a spacetime 
has been recently argued to emerge naturally \cite{TriantapEJTP}.
Namely, application of the
 optimal connectivity principle on a fuzzy version of node multisets can lead to a natural emergence of spacetime
in a quantum-mechanical setting,
where particles correspond to vacancies. It can be argued that nature prefers to form node configurations with
a finite number of nearest neighbors for each node, something which leads us away from 
the continuum as was recently shown \cite{TriantapICNFP15}, 
which might be argued to approach explicitly
old ideas on discreteness, some of which are also contained in \cite{Tegmark}. 

This picture might also resolve quantum-mechanical non-locality issues, 
since missing connections
 between defects with a common origin
spatially receding from each other, which in a dual picture 
 gives rise to quantum entanglement, explains the long-range correlations needed
to interpret EPR-type phenomena. It might thus assist in providing a 
precise implementation framework for techniques used when studying complexity
\cite{complexity} and quantum information \cite{Licata}.
Furthermore, it would be interesting to study possible connections of this setting with the string  
or the d-brane paradigm along the lines of \cite{Hooft2} or with the philosophy of causal sets \cite{causalsets}
which is generally compatible with the one presented here.

In order to study this system, 
 the $q$$=$$1$ 
Potts Hamiltonian is introduced:
\begin{equation} 
H_P= -\lambda\sum_{<i,j>}\delta(s_{i},s_{j})
\end{equation}
 \noindent which is usually employed in order
 to analyze disorder-order phenomena like percolation, spin-glass transitions, and even 
cognition modeling,
 the sum over the nodes $i, j$  being over nearest neighbors, $s_{i,j}=0{\rm ~or~} 1$, $\lambda>0$ the 
coupling strength and $\delta(s_{i},s_{j})$$=$$1$ when $s_{i}$$=$$s_{j}$$=$$1$, being
 zero otherwise. 
 Using the results in \cite{Conway}, it can be argued that this leads to the emergence of
a lattice $L_{G}$
 based on the roots of 
\begin{equation}
G\equiv E^{}_8 \times E_{8}^{\prime}
\end{equation}
\noindent  at the beginning of our Universe
\cite{TriantapEJTP}, 
as a ``liquid-to-solid" (freezing) ,
``glass-to-crystal" transition, higher dimensions leading to configurations bearing no relation to 
Lie-group symmetries.

Starting from an action $S_{lat}$ 
describing phenomena  like the emergence of $L_{G({\mathbb{C}})}$, 
assumed to belong to the
same universality class as $H_P$ does, which is given by
\begin{equation}
S_{lat}=\sum_{<i,j>}{\cal E}_{ij}\bar{\Psi}_{i}\Psi_{j},
\end{equation}
we embed $L_G$ in Euclidean ${\mathbb{C}}^{d}$ space
suitable for longer wavelengths, which leads to an action $S_{f}$ 
\noindent over a compact toric K${\rm {\ddot a}}$hler manifold expressed as
\begin{equation}
T_{G(\mathbb{C})}^{d}\equiv {\mathbb{C}}^{d}/L_{G({\mathbb{C}})}
\end{equation}
possessing a complex Lie-group structure \cite{TriantapICNFP15}, extending thus
the techniques used in \cite{Wetterich}:
\begin{eqnarray}
S_{f}(\Psi)&=&\int_{T_{G(\mathbb{C})}^{d}}
d^{d}x\det~\Biggl(\frac{i}{2}\bar{ \Psi}{\gamma^{m}\partial_{\mu}}\Psi+{\rm h.c.}\Biggl)~ \equiv \int_{T_{G(\mathbb{C})}^{d}}
d^{d}x\det{({\tilde E^m_{\mu}})} \nonumber \\ &=&
\int_{T_{G(\mathbb{C})}^{d}}d^{d}x\det{(E^m_{\mu})}\sum^{\infty}_{N=0}
\frac{\left[
-\sum\limits^{\infty}_{M=1}
\frac{{\rm tr}~\bigl(\left(\delta^{n}_{\nu}-(E^{-1}{\tilde E})_{\nu}^n\right)^M\bigr)}{M}
\right]^{N}}{N!}
\label{action}
\end{eqnarray}

\noindent with $d$=16 complex dimensions, $\gamma^{m}$ appropriate Dirac matrices, 
\begin{equation}
E^m_{\mu}\equiv<{\tilde E^m_{\mu}}>\equiv <\frac{i}{2}\bar{ \Psi}{\gamma^{m}\partial_{\mu}}\Psi+{\rm h.c.}>,
\end{equation}
\noindent lower-case
Greek and Latin indices corresponding to space-time and to the ``internal" Lorentz symmetry respectively, 
\begin{equation}
{\tilde N}\equiv MN\leq d
\end{equation}
 due to the Cayley-Hamilton theorem, and
integration over the torus $T_{G(\mathbb{C})}^{d}$ implying a UV cut-off 
$M_{{\rm Pl}}$$\sim$$L^{-1}_{{\rm Planck}}$
due to a minimal distance between nodes.  
The variables ${\bar \Psi}$, $\Psi$ 
 correspond in a dual sense to lattice vacancies and consist of two Weyl spinors of opposite chirality
following the Poincar${\rm \acute{e}}$-Hopf index theorem,
transforming like $\textbf{(248,1)}$ (SM fermions)
and $\textbf{(1,248)}$ (katoptrons) under $E^{}_8$$\times$$E_8^{\prime}$. 

Symmetry breaking of 
\begin{equation}
G\rightarrow H\equiv E^{}_7\times SL(2,\mathbb{C})\times E_7^{\prime}\times SL(2,\mathbb{C})^{\prime}
\end{equation}
\noindent via non-zero vevs
of antisymmetric bi-fermion operators transforming like ${\bf 248}$$\times$${\bf 248}$$\rightarrow$${\bf 248_a}$
under each of the 2 $E_{8}$s
yields a 
4
real-$d$ action $S_{eff}(E^m_{\mu},\psi,A_{\mu},\phi_{{\rm infl}})$ 
containing Einstein-Hilbert and Yang-Mills terms
in the approximation  
\begin{equation}
{\tilde E^m_{\mu}}-E^m_{\mu}\equiv
i\bar{\psi}\gamma^m\partial_{\mu}\psi \ll E^m_{\mu},
\end{equation}
with $A_{\mu}$ and $\phi_{{\rm infl}}$ Kaluza-Klein composite  
$E^{}_7\times E_7^{\prime}$-gauge and inflaton fields respectively,
 leading to the usual least-action-principle Euler-Lagrange equations in
Minkowski's
space-time. This happens after
 the $SL(2,\mathbb{C})$ subgroups of the two $E_{8}$s
break to their diagonal subgroup $S$$O$$(1$$,$$3$$)_D$ 
due to $M_{Pl}$-scale vevs of 2${\tilde N}$-fermion 
 operators (of order ${\mathcal O}(p/M_{\rm Pl})^{\tilde N}$
after Fourier transformation)
 for ${\tilde N}$$=$$2$ 
in Eq.(\ref{action}) coupling the left- and right-handed sectors and transforming like
(${\bf 1}$,${\bf 3}$,${\bf 1}$,${\bf 3}$) under $H$. 

At slightly lower energies, after the self-breaking
 of the SM-fermion $SU(3)$ family group due to a parity-violating $L_G$ asymmetry
which leaves the katoptron-generation symmetry intact and to which the
parity asymmetry of the SM can be traced originally,
similar operators transforming like 
(${\bf 24}$,${\bf 24}$)
 under 
\begin{equation}
{\tilde H}\equiv SU(5)\times SU(5)^{\prime}\subset E^{}_7\times E^{\prime}_7\subset H
\end{equation}
 lead to the breaking of ${\tilde H}$ to its diagonal subgroup
${\tilde H}$$\rightarrow$$SU(5)_{D}$, 
obviating the need for both outer automorphisms in \cite{TriantapEJTP}, in order to couple 
the SM-fermion and katoptron sectors of the theory with the same gauge groups apart from the 
katoptron generation group. 
Further symmetry breaking takes place starting from
 a unified critical coupling at $M_{{\rm Pl}}$, estimated via Schwinger-Dyson equations \cite{Miransky} as
 \begin{equation} 
\alpha({\rm M_{Pl}}) \sim 1/C_2(E_8) \sim 0.03
\end{equation}
down to the SM
having the known fermion-family structure with a non-perturbative BEH mechanism based on katoptrons close to the electroweak symmetry-breaking scale
\begin{equation}
\Lambda_{K}\sim M_{{\rm Pl}}\exp^{}\Bigl(-1.23C_2(E_8)\Bigr)\sim 1 {\rm ~TeV}
\end{equation}
\noindent  \cite{TriantapEJTP}
 in a way similar to QCD asymptotic freedom, due to the strongly-coupled katoptron-generation symmetry $SU(3)_{K}$. 
The katoptron Lagrangian of Section 2 is thus recovered.

It might
be  reminded at this point
 that the strongly-coupled source of electroweak symmetry breaking is 
expected to influence among others the order of the relevant phase transition which is crucial for 
electroweak baryogenesis \cite{Appelquist}. Moreover, katoptrons are expected to decay so fast due to the breaking
of their gauged generation symmetry that no problems related to Big-Bang nucleosynthesis are expected to arise.
Indeed, regarding a related issue, it has been noted 
elsewhere that, due to the absence of stable particles in the katoptron sector, the origins of Dark Matter in the present scenario
have to be traced to other solutions not based on particles but possibly on an
alternative spacetime topology instead \cite{TriantapEJTP}.

To conclude, an effort is made in this work 
 to study qualitatively the mirror-meson spectrum and decays resulting from katoptron theory,
 in order to lay the ground for more detailed 
and precise future relevant theoretical studies and computer simulations. On the 
phenomenological front, it has been argued previously that
the katoptron model, apart from the prediction of proton decay implied by gauge-coupling unification considerations,
 is expected to lead amongst others to deviations of the CKM-matrix element $|V_{tb}|$, of
third-generation SM-fermion weak couplings from 
their SM values due to the large mixing of heavier fermions with their mirror partners \cite{TriantapIJMPA} and 
of the muon magnetic moment from its expected value similarly to \cite{muon}.
Furthermore, it might lead to deviations connected to the decays of B mesons \cite{TriantapIJMPA}.

However, in case hadronic colliders lack the necessary resolution to detect such deviations
of higher-order quantum mechanical origin, the results presented here  indicate that 
LHC attention should be focused on the
enhancement of top-antitop pair production and acollinear top-antitop and bottom-antibottom
jets within specific invariant-mass bins as some of the most promising ways to find signals of
the various mirror mesons predicted in this framework. Discovering two distinct groups of spin-0 and spin-1 resonances
not only with masses separated roughly by a factor of six but also having the quantum-number assignments
listed in the present work would be very encouraging for the katoptron model. 
In parallel, confirmation of this picture would underline the 
need for a new 
high-energy (3-4 TeV) leptonic (possibly muonic) collider able to measure amongst others
 the left-right asymmetries
predicted by the special chiral character of mirror fermions \cite{TriantapIJMPA}.


\end{document}